\documentclass[12pt,preprint]{aastex}

\slugcomment{.}

\shorttitle{The Highly  Energetic Expansion of SN\,2010bh}
\shortauthors{Bufano et al.}

\begin{document}

\title{The Highly  Energetic Expansion of SN\,2010bh Associated with GRB\,100316D}

\author{Filomena Bufano\altaffilmark{1}, Elena Pian\altaffilmark{2,3,4}, Jesper Sollerman\altaffilmark{5}, 
Stefano Benetti\altaffilmark{6}, Giuliano Pignata\altaffilmark{7},  Stefano Valenti\altaffilmark{6},
Stefano Covino\altaffilmark{8},  Paolo D'Avanzo\altaffilmark{8}, Daniele Malesani\altaffilmark{9},
 Enrico Cappellaro\altaffilmark{6}, Massimo Della Valle\altaffilmark{10,11}, Johan Fynbo\altaffilmark{9},
 Jens Hjorth\altaffilmark{9}, Paolo A. Mazzali\altaffilmark{6,12}, Daniel E. Reichart\altaffilmark{13},
 Rhaana L. C. Starling\altaffilmark{14}, Massimo Turatto\altaffilmark{2}, Susanna D. Vergani\altaffilmark{8},
  Klass Wiersema\altaffilmark{14}, Lorenzo Amati\altaffilmark{15}, David Bersier\altaffilmark{16}, Sergio Campana\altaffilmark{8},
  Zach Cano\altaffilmark{16}, Alberto J. Castro-Tirado\altaffilmark{17}, Guido Chincarini\altaffilmark{18}, Valerio D'Elia\altaffilmark{19, 20},
 Antonio de Ugarte Postigo\altaffilmark{9},  Jinsong Deng\altaffilmark{21}, Patrizia Ferrero\altaffilmark{22},
 Alexei V. Filippenko\altaffilmark{23}, Paolo Goldoni\altaffilmark{24, 25}, Javier Gorosabel\altaffilmark{17},
 Jochen Greiner\altaffilmark{26},  Francois Hammer\altaffilmark{27}, Pall Jakobsson\altaffilmark{28},
 Lex Kaper\altaffilmark{29}, Koji S. Kawabata\altaffilmark{30}, Sylvio Klose\altaffilmark{31},
 Andrew J. Levan\altaffilmark{32}, Keiichi Maeda\altaffilmark{33}, Nicola Masetti\altaffilmark{34},
 Bo Milvang-Jensen\altaffilmark{9},  Felix I. Mirabel\altaffilmark{25, 35}, Palle M\o ller\altaffilmark{36},
 Ken'ichi Nomoto\altaffilmark{33}, Eliana Palazzi\altaffilmark{34}, Silvia Piranomonte\altaffilmark{19},
Ruben Salvaterra\altaffilmark{37},  Giulia Stratta\altaffilmark{19}, Gianpiero Tagliaferri\altaffilmark{8},
Masaomi Tanaka\altaffilmark{33}, Nial R. Tanvir\altaffilmark{14}, and Ralph A.M.J. Wijers\altaffilmark{29}
}

\altaffiltext{1}{INAF Post-Doc Fellow; INAF - Osservatorio Astronomico di Catania, Catania, Italy, 95123}
\altaffiltext{2}{INAF - Osservatorio Astronomico di Trieste, Via G.B. Tiepolo 11, I-34143, Trieste, Italy}
\altaffiltext{3}{Scuola Normale Superiore di Pisa, Piazza dei Cavalieri 7, 56126 Pisa, Italy}
\altaffiltext{4}{INFN - Sezione di Pisa, Largo Pontecorvo 3, 56127 Pisa, Italy}
\altaffiltext{5}{The Oskar Klein Centre, Department of Astronomy, AlbaNova, SE-106 91, Stockholm, Sweden}
\altaffiltext{6}{INAF - Osservatorio Astronomico di Padova, Vicolo dell'Osservatorio 5, I-35122, Padova, Italy}
\altaffiltext{7}{Departamento de Ciencias Fisicas, Universidad Andres Bello, Av. Republica 252, Santiago, Chile}
\altaffiltext{8}{INAF - Osservatorio Astronomico di Brera, Via Emilio Bianchi 46, Merate, I-23807, Italy}
\altaffiltext{9}{Dark Cosmology Centre, Niels Bohr Institute, University of Copenhagen, Juliane Maries Vej 30, DK-2100, Copenhagen, Denmark}
\altaffiltext{10}{INAF - Osservatorio Astronomico di Capodimonte, Salita Moiariello, 16, I-8013, Napoli, Italy}
\altaffiltext{11}{International Center for Relativistic Astrophysics Network, Pescara, Italy}
\altaffiltext{12}{Max-Planck Institut fur Astrophysik, Karl-Schwarzschildstr. 1, D-85748 Garching, Germany}
\altaffiltext{13}{University of North Carolina at Chapel Hill, Campus Box 3255, Chapel Hill, NC 27599-3255, USA}
\altaffiltext{14}{Department of Physics and Astronomy, University of Leicester, University Road, Leicester LE1 7RH, UK}
\altaffiltext{15}{INAF- Istituto di Astrofisica Spaziale e Fisica cosmica, Via Gobetti 101,  I-40129 Bologna, Italy}
\altaffiltext{16}{Astrophysics Research Institute, Liverpool John Moores University, 2 Rodney St, Liverpool, L3 5UX, UK}
\altaffiltext{17}{Instituto de Astrofisica de Andalucia (IAA-CSIC), Glorieta de la Astronomia s/n, 18008 Granada, Spain }
\altaffiltext{18}{Univerisit‡ Milano Bicocca, Dip. Fisica G. Occhialini, P.zza della Scienza 3, Milano 20126, Italy}
\altaffiltext{19}{INAF - Osservatorio Astronomico di Roma, via di Frascati 33, 00040 Monte Porzio Catone, Rome, Italy}
\altaffiltext{20}{ASI-Science Data Center, Via Galileo Galilei, I-00044, Frascati, Italy}
\altaffiltext{21}{National Astronomical Observatories, CAS, 20A Datun Road, Chaoyang District, Beijing 100012, China}
\altaffiltext{22}{Instituto de Astrof'sica de Canarias (IAC), 38200 La Laguna, Tenerife, Spain}
\altaffiltext{23}{Department of Astronomy, University of California, Berkeley, CA 94720-3411, USA}
\altaffiltext{24}{Laboratoire Astroparticule et Cosmologie, 10 rue A. Domon et L. Duquet, 75205 Paris Cedex 13, France}
\altaffiltext{25}{Service d'Astrophysique, DSM/IRFU/SAp, CEA-Saclay, 91191 Gif-sur-Yvette, France}
\altaffiltext{26}{Max-Planck Institut f\"ur extraterrestrische Physik, Giessenbachstrasse 1, D-85740 Garching, Germany }
\altaffiltext{27}{GEPI-Observatoire de Paris Meudon. 5 Place Jules Jannsen, F-92195, Meudon, France}
\altaffiltext{28}{Centre for Astrophysics and Cosmology, Science Institute, University of Iceland, Dunhagi 5, 107 Reykjav\'ik, Iceland}
\altaffiltext{29}{Astronomical Institute Anton Pannekoek, University of Amsterdam, Science Park 904, 1098 XH Amsterdam, The Netherlands}
\altaffiltext{30}{Hiroshima Astrophysical Science Center, Hiroshima University,  1-3-1 Kagamiyama, Higashi-Hiroshima, Hiroshima 739-8526, Japan}
\altaffiltext{31}{Thuringer Landessternwarte Tautenburg, Sternwarte 5, D-07778 Tautenburg, Germany}
\altaffiltext{32}{Department of Physics, University of Warwick, Coventry CV4 7AL, UK}
\altaffiltext{33}{Institute for the Physics and Mathematics of the Universe,   University of Tokyo,   5-1-5 Kashiwanoha, Kashiwa, Chiba 277-8583, Japan}
\altaffiltext{34}{INAF - Istituto di Astrofisica Spaziale e Fisica cosmica, Via Gobetti 101, I-40129 Bologna, Italy}
\altaffiltext{35}{IAFE-CONICET-UBA. cc67, suc 28, Buenos Aires, Argentina}
\altaffiltext{36}{European Organization for Astronomical Research in the Southern Hemisphere (ESO), Karl-Schwarzschild-Str. 2, 85748 Garching,  Germany}
\altaffiltext{37}{Dipartimento di Fisica e Matematica, Universit\`a dell'Insubria, via Valleggio 7, 22100 Como, Italy}

\begin{abstract}

We present the spectroscopic and photometric evolution of the  nearby ($z = 0.059$) spectroscopically confirmed type Ic
supernova, SN\,2010bh, associated with the soft, long-duration gamma-ray burst (X-ray flash) GRB\,100316D.
Intensive follow-up observations of SN\,2010bh were performed at the ESO Very Large Telescope (VLT) using the X-shooter and FORS2 instruments. 
Owing to the detailed temporal coverage and the extended wavelength range (3000--24800 \AA), we obtained  an unprecedentedly rich spectral sequence  among the hypernovae, making SN\,2010bh one of the best studied representatives of this SN class. We find that  SN\,2010bh  has a more rapid rise to maximum brightness (8.0 $\pm$ 1.0 rest-frame days) and a fainter absolute peak luminosity ($L_{\rm bol}\approx 3 \times 10^{42}\,erg\,s^{-1}$) than previously observed SN events associated with GRBs.  
Our estimate of the ejected $^{56}$Ni mass is $0.12 \pm 0.02$ M$_\odot$. 
From the broad spectral features we measure expansion velocities up to 47,000\,km\,s$^{-1}$, higher than those of SNe 1998bw (GRB\,980425) and 2006aj (GRB\,060218). 
Helium absorption lines He\,I $\lambda$5876 and He\,I 1.083 $\mu$m, blueshifted by $\sim$20,000--30,000 km\,s$^{-1}$ and $\sim$28,000--38,000 km\,s$^{-1}$, respectively, may be present in the optical spectra. However, the lack of coverage of the He\,I  2.058$\mu$m line prevents us from confirming such identifications. The nebular spectrum, taken at $\sim$186 days after the explosion, shows a broad 
but faint [O\,I] emission at 6340\AA.
The light-curve shape and  photospheric expansion velocities of SN\,2010bh suggest that we witnessed a highly energetic explosion with a small ejected mass ($E_{\rm k} \approx 10^{52}$ erg and $M_{\rm ej} \approx 3$~M$_\odot$). The observed properties of SN\,2010bh further extend the heterogeneity of the class of GRB supernovae.

\end{abstract}



\section{Introduction}
During the past decade, the link between long-duration gamma-ray bursts  (GRBs) and type Ic core-collapse supernovae (SNe; e.g., \citealt{Filippenko97}) has been firmly established; see \citet{Woosley} and \citet{HjoBlo} for reviews. The first clear case occurred in 1998, when the luminous SN\,1998bw, at  $z = 0.0085$, was found spatially and temporally coincident with GRB\,980425 \citep{Galama98bw}.  
The GRB-SN connection was supported in 2003 by two further associations  between nearby GRBs and spectroscopically confirmed SNe: GRB\,030329/SN\,2003dh at redshift $z = 0.17$ (\citealt{Hjorth}; \citealt{Stanek};  \citealt{Matheson03dh}) and GRB\,031203/ SN\,2003lw at $z = 0.11$ (\citealt{Malesani};  \citealt{Thomsen03lw}; \citealt{Gal-Yam03lw}; \citealt{cobb04}). 
The most recent case of a spectroscopic connection is GRB\,060218/SN\,2006aj ($z = 0.033$, \citealt{Campana}; \citealt{Pian06aj}; \citealt{Mirabal06aj}; \citealt{Modjaz06aj}; \citealt{Cobb06aj}; \citealt{Ferrero}). 
On average, SNe associated with classical GRBs appear to be more luminous at peak than SNe~Ic not accompanied by GRBs, while SNe associated with X-ray flashes have maximum luminosities more similar to those of normal SNe~Ic (see, e.g., \citealt{Pian06aj}; \citealt{Pignata09bb}; \citealt{Drout}).  However, SNe associated with both GRBs and X-ray flashes exhibit broader features in their spectra, indicating unusually large expansion velocities. 
From the modeling of their light curves and spectra, very high explosion energies are inferred ($\sim10^{52}$ erg,  about 10 times higher
than typical SNe), which made earn them  the name of hypernovae (HNe; \citealt{Paczynski};  \citealt{Iwamoto98bw}).
The GRB-SN connection has been best studied  at low redshift ($z < 0.2$), where the clear, spectroscopically confirmed cases have been detected.
Although GRBs at these low redshifts are rarely observed, the association between GRBs and SNe has been extended up to $z\sim1$, (corresponding to a look back time of about 60\% the age of the Universe) in a number of GRB-SNe which have been identified through single epoch spectra characterized by the presence of SN features (\citealt{Lazzati}; Della Valle et al. 2003, 2006, 2008; \citealt{Soderberg05}; \citealt{Bersier};  \citealt{Cobb09nz}; \citealt{Cano0.5}; \citealt{Sparre}; \citealt{Berger11}).   
The investigation of all confirmed GRB-SN associations is critical to understand the  nature of their progenitors and the mechanism by which powerful stripped-envelope SNe produce ultra-relativistic jets (e.g., \citealt{Zhang}; \citealt{Uzd}; \citealt{Mazzali08d}; \citealt{Fryer}; \citealt{Lyutikov}).
 
GRB\,100316D \citep{BATtrigger}  is a low-redshift event ($z = 0.059$; \citealt{Vergani}; \citealt{Starling}; see also \S \ref{host}), whose  prompt emission is  characterized by a very soft spectral peak, similar to that of X-ray flashes (XRFs; GRBs with energy peak at low frequency; \citealt{Heise}),  and a slowly decaying flux. A few days after its detection, an associated type Ic supernova was identified through the spectral features of the early optical counterpart: SN\,2010bh  (Chornock et al. 2010a,b; \citealt{Wiersema}; \citealt{BufanoCBET}).
Similar to the low-redshift X-ray flash GRB\,060218, GRB\,100316D had an unusually long duration ($T_{90} > 1300$\,s) and a spectral-hardness evolution with a stable and soft spectral shape throughout the prompt and late-time emission \citep{Starling}. 
The early X-ray spectrum of GRB\,100316D, like that of GRB\,060218 \citep{Campana}, is best described by a power law plus a thermal component  (\citealt{Starling}; \citealt{Fan}). 
The latter may be  the signature of either the shock breakout following core collapse (\citealt{Campana}; \citealt{Waxman}), or  additional radiation from the central engine (e.g., \citealt{Ghisellini}; \citealt{Li}; \citealt{Chevalier}).  

We intensively monitored  the optical and near-infrared (NIR) spectrophotometric evolution of SN\,2010bh, starting $\sim$12\,hr after the {\it Swift}/BAT GRB trigger, which  occurred on  2010 March 16 at 12:44:50  (\citealt{BATtrigger}; UT dates are used throughout this paper), until  about 2 months past the discovery, when solar constraints made no longer observable the SN.   To enable accurate host-galaxy subtraction, we reobserved the field 0.5--1 yr after the explosion.   This campaign was the outcome of the coordination of various  observing programs at the European Southern Observatory (ESO, Chile) Very Large Telescope (VLT)\footnote{VLT observations were taken within the GTO programs 084.D-0265 and 085.D-0701 (P.I. S. Benetti) and 084.A-0260 and 085.A-0009 (P.I. J. Fynbo) at UT2/X-shooter, and GO program 085.D-0243 (P.I. E. Pian) at UT1/FORS2.  SN\,2010bh photometry on March 23 and 28 was obtained with GO program 084.D-0939 with UT1/FORS2 (P.I. K. Wiersema).  Observations were performed in ToO mode.} and at the Cerro Tololo Interamerican Observatory (CTIO, Chile). The  spectral behavior of SN\,2010bh from day 1.39 to day 21.2 has also been discussed by \citet{Chornock} and photometry during the first 3 months after explosion has been presented by \citet{Cano} and \citet{Olivares}.  
The larger wavelength range and improved phase coverage of  our observations allow us to analyze the early phases in more detail and to push the investigation into the late evolutionary stages.

In \S \ref{data}, we present the  dataset and describe the data reduction methods,  while in \S \ref{results} we show the 
spectrophotometric evolution of SN\,2010bh and compare it to that of the previously well-studied GRB-SNe.
In \S \ref{discussion}, we discuss the derived properties of the progenitor star and the explosion parameters, and we summarize our conclusions.

\section{Data Acquisition and Reduction} \label{data}

\subsection{Photometry}

$UBVRI$ photometry of  SN\,2010bh was obtained with the FOcal Reducer and low dispersion Spectrograph (FORS2; field of view [FOV] $6.8' \times 6.8'$; scale $0.25''$ pixel$^{-1}$; \citealt{FORS}) at the ESO VLT UT1  and with the 0.41-m Panchromatic Robotic Optical Monitoring and Polarimetry Telescopes (PROMPTs) 1 and 5 located at CTIO (FOV $10' \times 10'$, $0.6''$ pixel$^{-1}$; \citealt{Prompt}). The $R$-band images from the X-shooter Acquisition Camera (FOV $1.47' \times 1.47'$; $0.173''$ pixel$^{-1}$; \citealt{Dodorico}) were also used to cover the early evolution. 
The data were reduced with standard techniques using IRAF \footnote{IRAF is distributed by the National Optical Astronomy Observatories, which are operated by the Association of Universities for Research in Astronomy, Inc. under contract with the National Science Foundation.} tasks.
Since the SN exploded in a region with a complex background, a template subtraction method,  based on the ISIS package (\citealt{Alard98}, \citealt{Alard00}),  was applied to remove contamination from the host-galaxy light. As no pre-explosion observations were available of SN\,2010bh, we used images obtained with VLT/FORS2 on September 17 (about 185 days past explosion; see Table \ref{templates}),  assuming that the SN flux contribution was negligible at this epoch (see \S \ref{results}). For PROMPTs observations, template images were acquired on 2011 February 2, with the exception of the $I$ band at PROMPT5, for which the template image was acquired on  2011 January 26 (see Table \ref{templates}).  

\begin{figure}
 \begin{center}
 \includegraphics[scale=.7]{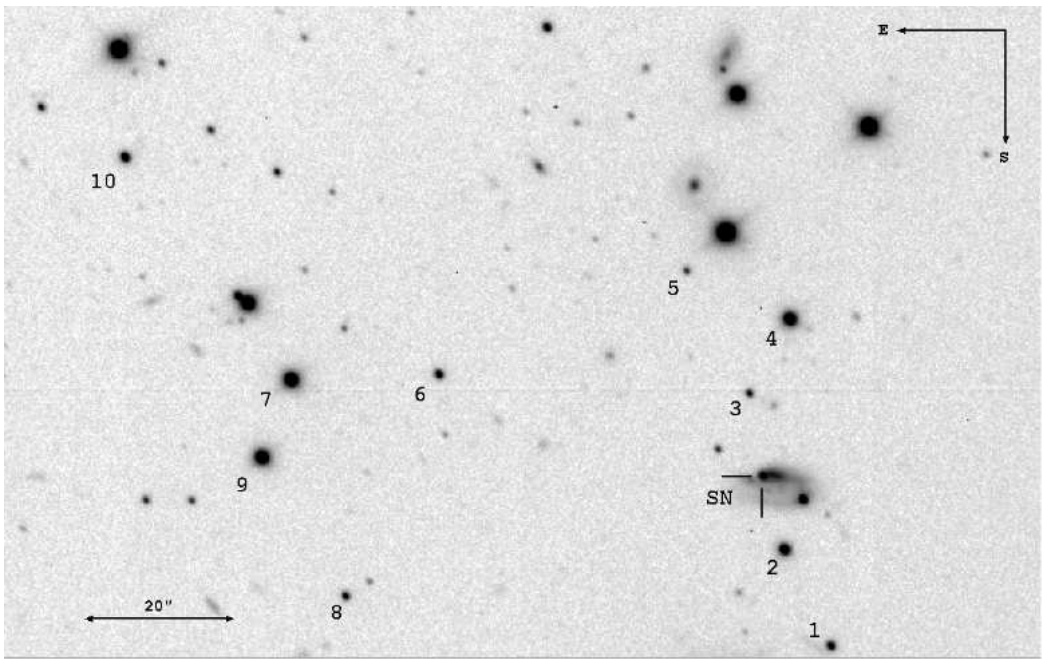}
 \caption{Close-up view of SN\,2010bh field from the $R$-band image taken with VLT/FORS2  on  2010 April 3 (scale in the lower-left corner). 
 SN\,2010bh and the sequence of local reference stars (Table \ref{SeqStars}) are reported.  \label{FC} }
 \end{center}
 \end{figure}

\begin{figure}
\includegraphics[scale=.8]{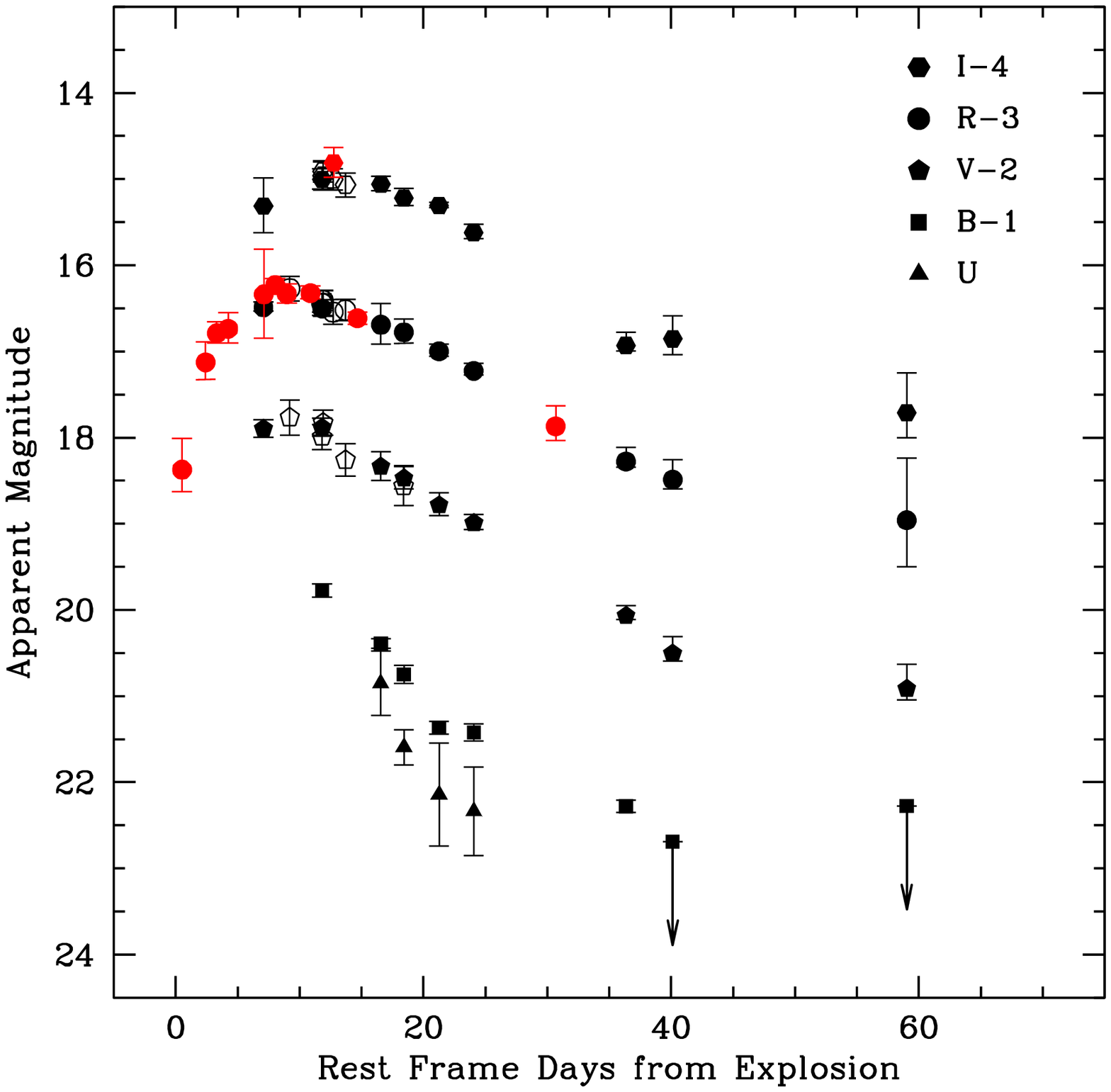}
\caption{$UBVRI$ light curves of SN\,2010bh.  The abscissa represents the rest-frame time after the explosion, which is assumed to be
coincident with the burst trigger (2010 March 16.53; \citealt{BATtrigger}). Magnitudes have not been corrected for Galactic and host galaxy extinction. K-corrections have been applied to the $BVRI$ magnitudes.  FORS2 and X-shooter $R$-band photometry is reported with black and red solid circles, respectively.  Open symbols are used for $VRI$ magnitudes obtained with PROMPT. For clarity, the light curves are vertically displaced by the amount reported in the legend for each filter.   \label{10bh_lc}}
\end{figure}

 A point-spread-function (PSF) fitting method was applied to measure the SN magnitudes in the difference images. 
 In particular, the PSF was derived from field stars measured on either the template or target image, whichever had the worst seeing. The kernel size was scaled according to the seeing to avoid wings truncation. We also performed aperture photometry on the subtracted images obtaining very similar results.  The uncertainties were estimated by means of artificial stars with the same  magnitude as the SN: we firstly placed them  close to the SN position (within few pixels) and then in the opposite side of the galaxy, where the background residuals were similar.  The two approaches, intended to test for the effect of different background  contributions, gave similar results. 
By observing several photometric standard fields, from \citet{Landolt} for the $U$ band and from \citet{Stetson} for the $BVRI$ filters, we obtained the color equations for each night and instrument, and used them to transform  instrumental magnitudes to the  standard photometric system.
We calibrated the magnitudes of a local  sequence of stars in the SN\,2010bh field  over four photometric nights (April 3, 5, 8 and 11) and used them to obtain the photometric zero-points for the non-photometric nights (Fig. \ref{FC} and Table \ref{SeqStars}). 
 Because of the significant color term, we applied  a calibration correction (S-correction) to the PROMPTs instrumental SN magnitudes
   to transform them into a standard photometric system following  \citet{Pignata08}.    
Finally, a K-correction based on the nearly simultaneous spectra was applied to the observed $BVRI$ magnitudes.
For the $U$-band magnitudes, the signal-to-noise (S/N) ratio was too low to measure a reliable K-correction  from the spectra. 
$UBVRI$ magnitudes of the supernova are reported in Table \ref{SN2010bh_phot}  and plotted in Figure \ref{10bh_lc}.
 We computed upper limit magnitudes (Table \ref{SN2010bh_phot} and Fig. \ref{10bh_lc}) placing an artificial star at the SN position on the background subtracted images and decreasing its magnitude down to the point of detectability over the background level. At template images epochs, limiting magnitudes were measured directly over the host galaxy level, reported in Table \ref{SN2010bh_phot} but not in  Fig. \ref{10bh_lc} because they do not set tight limits to the late phase decline.

\subsection{Spectroscopy}\label{Spec}

We followed the spectroscopic evolution of SN\,2010bh  with VLT/X-shooter  (12 epochs)  and VLT/FORS2 (8 epochs), as reported in Table \ref{SN2010bh_spec}. 
When possible, the spectra were acquired with the slit positioned along the North-South direction, to minimize the host-galaxy contamination. 
For both instruments, the effects of atmospheric dispersion \citep{Filippenko82} at high airmass have been reduced by using an atmospheric dispersion  corrector. Simultaneous UV, VIS, and NIR spectra ($\sim$ 3000--24800 \AA)  were taken with X-shooter using slit widths of 1.0\arcsec, 0.9\arcsec, and 0.9\arcsec for each arm, respectively \citep{Dodorico}.
We used a nodding throw along the slit (nodding lengths of 2\arcsec\, and  4\arcsec) to obtain better sky subtraction. 
The data were reduced using version 0.9.4 of the ESO X-shooter pipeline \citep{Goldoni} with the calibration frames (biases, darks, arc lamps, and flatfields)  taken during daytime.  After reducing the data using more advanced versions of the software, no relevant changes were found.
With FORS2 we used the 300V grism (3300--9000 \AA) and a slitwidth of 1\arcsec.

Both X-shooter and FORS2 spectra were extracted using standard IRAF tasks.
Spectrophotometric and telluric standard-star exposures taken on the same night as the SN\,2010bh observations  were used to flux-calibrate the extracted spectra and to remove telluric absorption features.
We checked the absolute flux calibration of the spectra by using the nearly simultaneous $R$-band magnitudes.
Figure \ref{10bh_spectevol} shows the spectral sequence after correcting for both Galactic and host galaxy reddening; both the wavelength scale and epochs are reported in the
host-galaxy rest frame (see \S \ref{host} and \S \ref{phot_spec}).    The most prominent emission lines of the host galaxy have been removed. Figures \ref{10bh_opt} is a zoom-in of the spectral sequence in the optical range.
The three spectra taken in the nebular phase (2010 September 28 to October 1; see Table \ref{SN2010bh_spec}) have been combined to improve the signal-to-noise ratio.  The coadded spectrum  is shown in  Figure \ref{late_spec} (see \S \ref{neb_spec}). 
 
\section{Results}\label{results}

\subsection{Host-Galaxy Properties}\label{host}

SN\,2010bh exploded at $\alpha = 07^{\rm h}10^{\rm m}30^{\rm s}.53$ and $\delta = -56^{\circ}15'19\farcs78$ (J2000; \citealt{Starling}) in a bright anonymous galaxy.   We measured the host-galaxy redshift by calculating the average shift of the central wavelengths of its strongest emission lines ([O\,II] $\lambda$3727, Ne\,III $\lambda$3869, [O\,III] $\lambda\lambda$4959, 5007, H\,I Balmer lines, He\,I $\lambda$5876,  [O\,I] $\lambda$6300,  [N\,II] $\lambda$6584, [Si\,II] $\lambda\lambda$6716, 6731) in each of the 12 X-shooter spectra and correcting it for the radial component of the Earth's heliocentric motion.  The weighted mean of the resulting heliocentric  redshifts  is $z = 0.0592 \pm 0.0001$. 
 The high precision of the redshift estimate was possible thanks to the accuracy of the wavelength solution over the whole wavelength range of the X-shooter spectra (2 km s$^{-1}$ for the UV and VIS arms). 
This value is in  good agreement with the values presented in previous works ($z \approx 0.059$, \citealt{Vergani}; $z=0.0591 \pm 0.0001$, \citealt{Starling}; $z=0.0593$, \citealt{Chornock}).
For a concordance cosmology  (Hubble constant $H_0 = 73$ \, km\,s$^{-1}$\,Mpc$^{-1}$,  $\Omega_{\Lambda} = 0.73$, and $\Omega_{m} = 0.27$), we obtained a luminosity distance of about 254 Mpc (i.e., distance modulus $\mu$= 37.02 mag).

\begin{figure}
\includegraphics[scale=0.8]{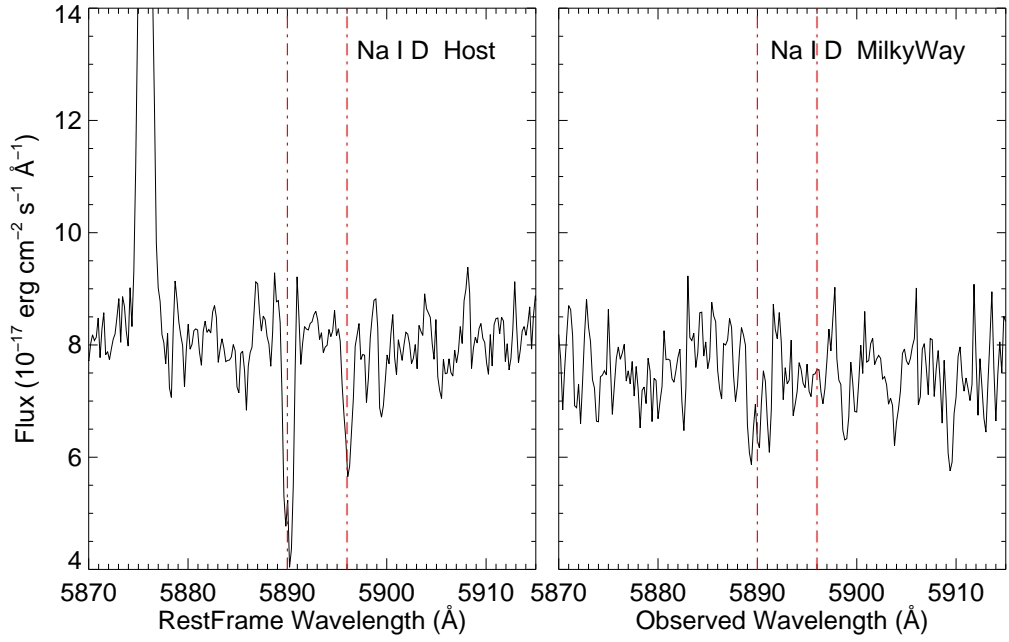}
\caption{Na\,I\,D doublet absorption  lines in the SN\,2010bh  spectrum obtained by averaging the early-time (phases 2.4d, 3.3d and 4.2d, see Table \ref{SN2010bh_spec}) X-shooter spectra. \label{redd}}
\end{figure}

We estimated the host-galaxy extinction by measuring the total equivalent width (EW)  of the interstellar Na\,I\,D absorption doublet ($\lambda\lambda$5890.0, 5895.9) with the  assumption of a gas-to-dust ratio similar to the average ratio in our Galaxy.
Measurements were performed on a spectrum obtained combining the almost featureless early-epoch X-shooter spectra (phases 2.4d, 3.3d and 4.2d, see Table \ref{SN2010bh_spec}). We  found  EW($\lambda$5890.0)$_{\rm host} = 0.59 \pm 0.05$ \AA\ and  EW($\lambda$5895.9)$_{\rm host} = 0.30 \pm 0.02$ \AA, giving a total EW(Na\,I\,D)$_{\rm host} = 0.89 \pm 0.07$ \AA\ (Fig. \ref{redd}). 
Applying the relation by \citet{TurattoEW}, $E(B-V)= 0.16 \times {\rm EW(Na\,I\,D)}$, we obtained $E(B-V)_{\rm host} = 0.14 \pm 0.01$ mag, which is the value we adopt throughout this work.
For the Milky Way  extinction, we measured EW($\lambda$5890.0)$_{\rm MW} = 0.40 \pm 0.10$ \AA, while the second doublet component was not detectable (Fig. \ref{redd}).
Assuming a flux ratio 2:1 between the two absorption lines, we obtained a total  EW(Na\,I\,D)$_{\rm MW} = 0.60 \pm 0.15$ \AA, implying $E(B-V)_{\rm MW} = 0.10 \pm 0.03$ mag.
 This value is in agreement with that found by \citet{Schlegel},  $E(B-V)_{\rm MW} = 0.12$ mag.
We decided to adopt  the latter  because of the large uncertainty in our estimate of EW(Na\,I\,D)$_{\rm MW}$.

Recently, \citet{Dovi}  and  \citet{OlivaresRed}  claimed that Na\,I\,D absorption may be a bad proxy for the extinction, especially if one uses low-resolution spectra where the two doublet lines cannot be resolved. Although we have a higher dispersion in the X-shooter spectra that allows us to separate the two components,  we checked our result by estimating the reddening  inside the host galaxy (along the line of sight) from the Balmer-line intensity ratios of the H\,II region coincident with the SN. 
Firstly, we corrected both X-shooter and FORS2 spectra for the Milky Way extinction, then, assuming Case B recombination ($T = 10^4$ K; \citealt{Oster}), we measured the H$\alpha$/H$\beta$ ratios from each spectrum. We obtained an average value of $E(B-V)_{\rm host} = 0.18 \pm 0.06$ mag,  in  agreement with that used in this work ($E(B-V)_{\rm host} = 0.14$ mag).\\
An independent and consistent estimate of the reddening has been given by \citet{Cano}, who found a host galaxy color excess $E(B-V)_{\rm host}=0.18 \pm 0.08$ mag comparing the SN\,2010bh colors with those of the type Ibc SN sample studied by \citet{Drout}.
A higher value for the  host galaxy reddening ($E(B-V)_{\rm host}=0.39 \pm 0.03$ mag) has been found by  \citet{Olivares}, by fitting a broad-band SED constructed  using GROND and {\it Swift}/XRT data. While \citet{Cano} and  \citet{Olivares} have estimated $E(B-V)_{\rm host}$ from indirect methods (statistics of Type  Ic SNe and SED modeling, respectively), our procedure is based on a direct estimate of the dust amount in the line of sight of the SN from the optical spectra, the only necessary underlying assumption consisting in considering the dust-to-gas ratio as constant and equal to the Galactic one.

We also used the 12 X-shooter spectra  to estimate the metallicity of the bright region underlying SN\,2010bh.
We measured both the N2 and O3N2 diagnostic ratios \citep{Pettini}, obtaining an average oxygen abundance 12 + log(O/H) = 8.20 $\pm$ 0.24, where the error is dominated by the uncertainties associated with the adopted linear relationships.
 The values of $\sim$8.2 reported by \citet{Chornock} at the SN location, which is based on  a spectrum at +3.3 days after the explosion,  and   8.2 $\pm$ 0.1  by \citet{Levesque}, which is based on a spectrum at +52 days, are in excellent agreement with our estimate. 
 From the spectrum of the H\,II region located  close to the SN, \citet{Starling} found an oxygen abundance of 8.23 $\pm$ 0.15.
 
\subsection{SN\,2010bh Light Curves}

Figure \ref{10bh_lc} illustrates the SN\,2010bh light curves.  Our $R$-band light curve traces well the early evolutionary stages; the SN  reaches maximum light  ($M_R \approx -18.5$ mag) at $8.0 \pm 1.0$  rest-frame days past the explosion, confirming the rise time independently found by \citet{Cano} and \citet{Olivares}. This is the  steepest rise to maximum brightness ever found among both SNe associated with GRBs and broad-lined (BL) SNe for which explosion dates have been well constrained (e.g., SN\,2002ap: $R_{\rm max}$ at 12 days; \citealt{Mazzali02ap}; \citealt{Foley02ap}; SN\,2003jd: $R_{\rm max}$ at $\sim 16$ days; \citealt{Valenti03jd}; SN\,2005nc: $R_{\rm max}$ at $\sim 12$ days; \citealt{DellaValle2006}).

\begin{figure}
\includegraphics[scale=0.7]{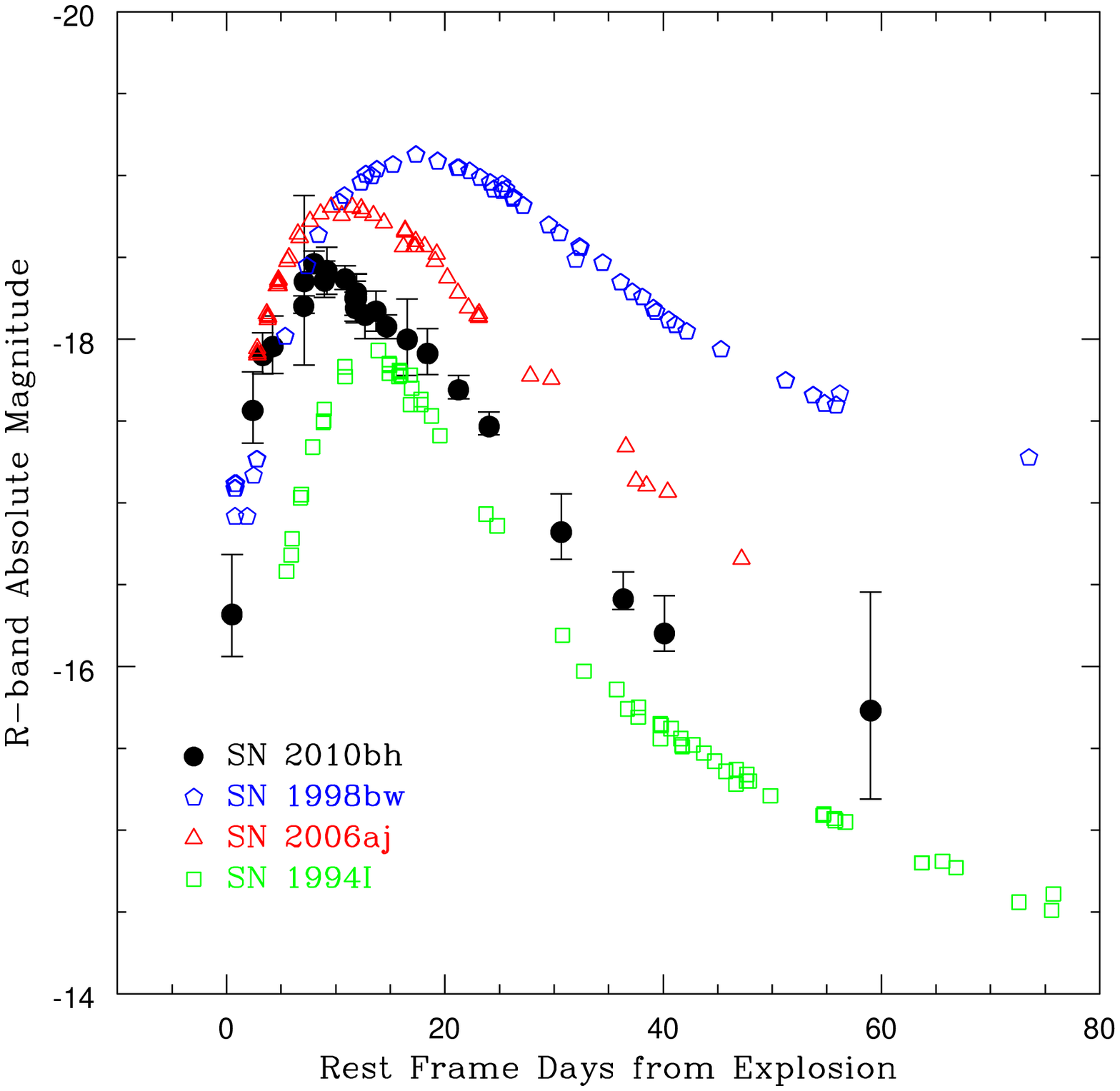}
\caption{Light curve of SN\,2010bh compared to those of SN\,1998bw 
(\citealt{Galama98bw}; \citealt{Patat98bw}) and SN\,2006aj (\citealt{Sollerman06aj}; \citealt{Pian06aj}; \citealt{Ferrero}), both of which were associated with a GRB/XRF, and to that of the type Ic SN\,1994I (\citealt{Richmond94I}), not accompanied by a high-energy event. All SN light curves have been corrected for total reddening along the line of sight and are reported in the host-galaxy rest frame. The epochs are given with respect to the burst detection. For SN\,1994I, the explosion date was obtained
from the light-curve models of \cite{Iwamoto94I}. \label{Ic_comp}}
\end{figure}

 A decline of 0.056 $\pm$ 0.015 mag day$^{-1}$ is measured between 0 and 15 days after $R$-band maximum.
Good sampling of the post-maximum phases was also obtained in the $UBVI$ bands (Fig. \ref{10bh_lc}).
In Figure \ref{Ic_comp}, we compare the $R$-band light curve of SN\,2010bh  with those of two previous well-sampled GRB-SNe: SN\,1998bw (\citealt{Galama98bw}; \citealt{Patat98bw}) and SN\,2006aj (\citealt{Sollerman06aj}; \citealt{Pian06aj}; \citealt{Ferrero}).   The light curve of a more typical  type Ic SN (SN\,1994I; \citealt{Richmond94I}) is also shown.

After maximum brightness, SN\,2010bh and SN\,1998bw have a similar behavior,  but the decay rate of SN\,2010bh between the last two $R$-band points corresponds to $\sim 0.03$ mag day$^{-1}$ in the rest frame, to be compared with that of SN\,1998bw (0.013 mag day$^{-1}$).   
Using the decline rates found by \citet{Patat98bw} for SN\,1998bw, we estimate the magnitudes of SN\,2010bh at the epoch when the VLT subtraction images were acquired (+166.5 rest-frame days from maximum; see Table \ref{templates}): $V \approx 25.0$, $R \approx 23.8$, and $I \approx 23.6$ mag. 
Considering this as an upper limit, if it was indeed the flux level of SN\,2010bh,  it would cause an oversubtraction and an underestimate of the SN fluxes in the $V$, $R$, and $I$ bands by less than 0.03 mag around maximum and 0.15 mag in the latest epochs.  
For each epoch, the corresponding inferred possible contamination by a residual SN flux has been included in the error estimate (Table  \ref{SN2010bh_phot}).

SN\,2010bh is less luminous than the other two GRB-SNe (Fig.\,\ref{Ic_comp}), which  suggests a smaller amount of ejected  $^{56}$Ni mass. 
Moreover, since the width of a light curve scales with the ratio between the total ejected mass $M_{\rm ej}$ and  the total explosion kinetic energy $E_{\rm k}$  (Arnett 1982, 1996), the fast evolution of SN\,2010bh likely reveals  a relatively highly energetic explosion and/or a small $M_{\rm ej}$.
 On the other hand, a highly asymmetric explosion may also result in a faster expansion in the polar direction, leading to a short diffusion time and a fast rise time \citep{Maeda2006}. 
 
We note that our $R$-band photometry (neither corrected for Galactic or host extinction) at 0.5 days  is $\sim$1 mag and $\sim$0.6 mag fainter than those of \citet{Cano} and  \citet{Olivares}, respectively.   This implies that we observe in our data the smooth rise in flux typical of a SN, and no evidence of the extra early component that they interpret as  shock breakout.  While we processed all raw data in a homogeneous way and consider this first flux point and its uncertainty formally correct, we caution that at these low flux levels the X-Shooter acquisition camera imaging data may depend very critically on the assumed background.

\subsection{Spectra in the Photospheric Phase}\label{phot_spec}

\begin{figure}
\includegraphics[scale=.7]{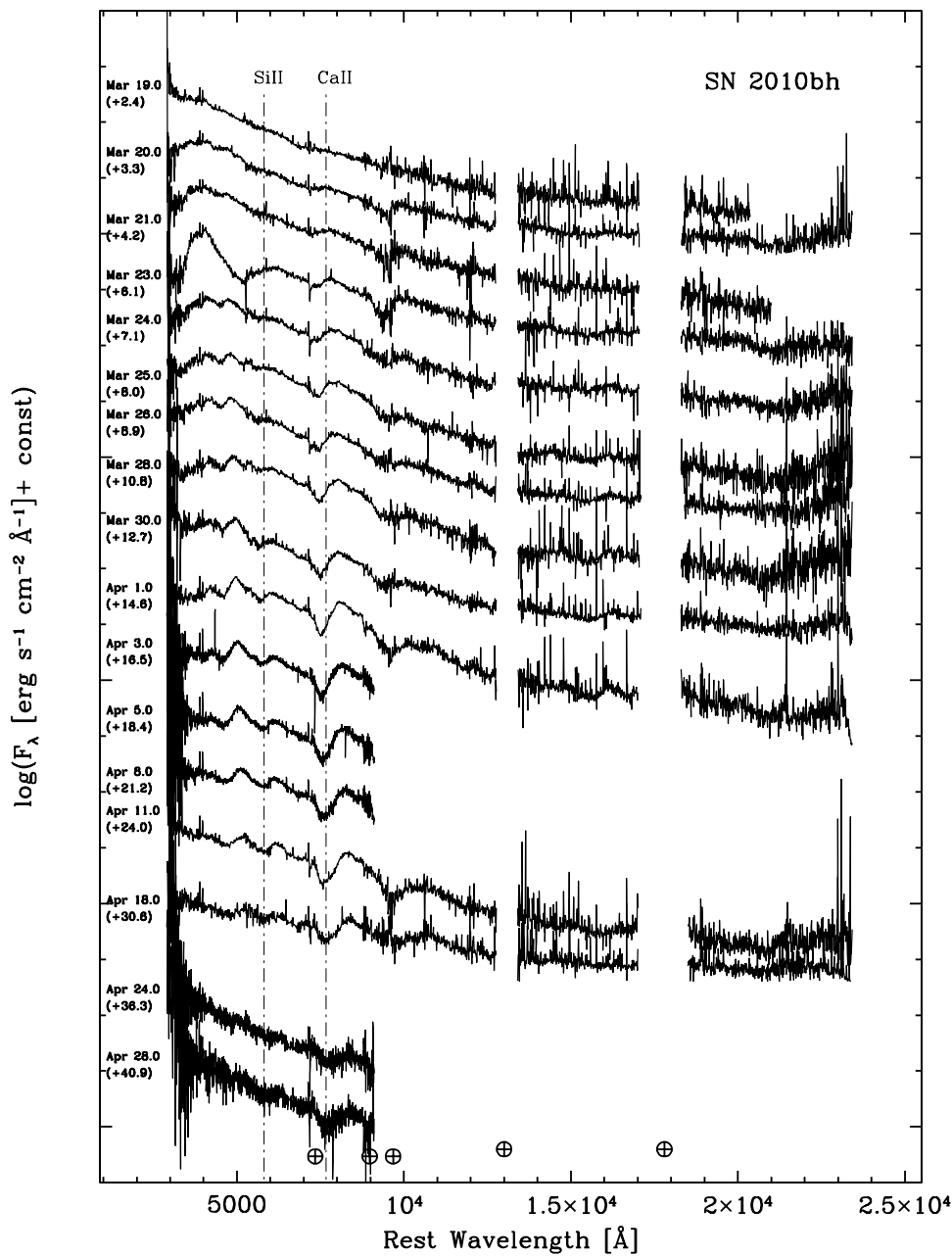}
\caption{SN\,2010bh spectral evolution. The phase is given in rest-frame days after  the explosion, assumed to be
coincident with the GRB start time (2010 March 16.53; \citealt{BATtrigger}).
The spectra are corrected for total (Milky Way + host-galaxy) reddening, shifted to the galaxy rest frame, 
vertically displaced and rebinned for clarity. 
Dot-dashed vertical lines indicate the wavelengths of the minima in the Si\,II and Ca\,II  absorption lines on the April 18 spectrum.  The most prominent emission lines of the host galaxy have been removed and telluric band positions indicated. \label{10bh_spectevol}}
\end{figure}

\begin{figure}
\includegraphics[scale=.7]{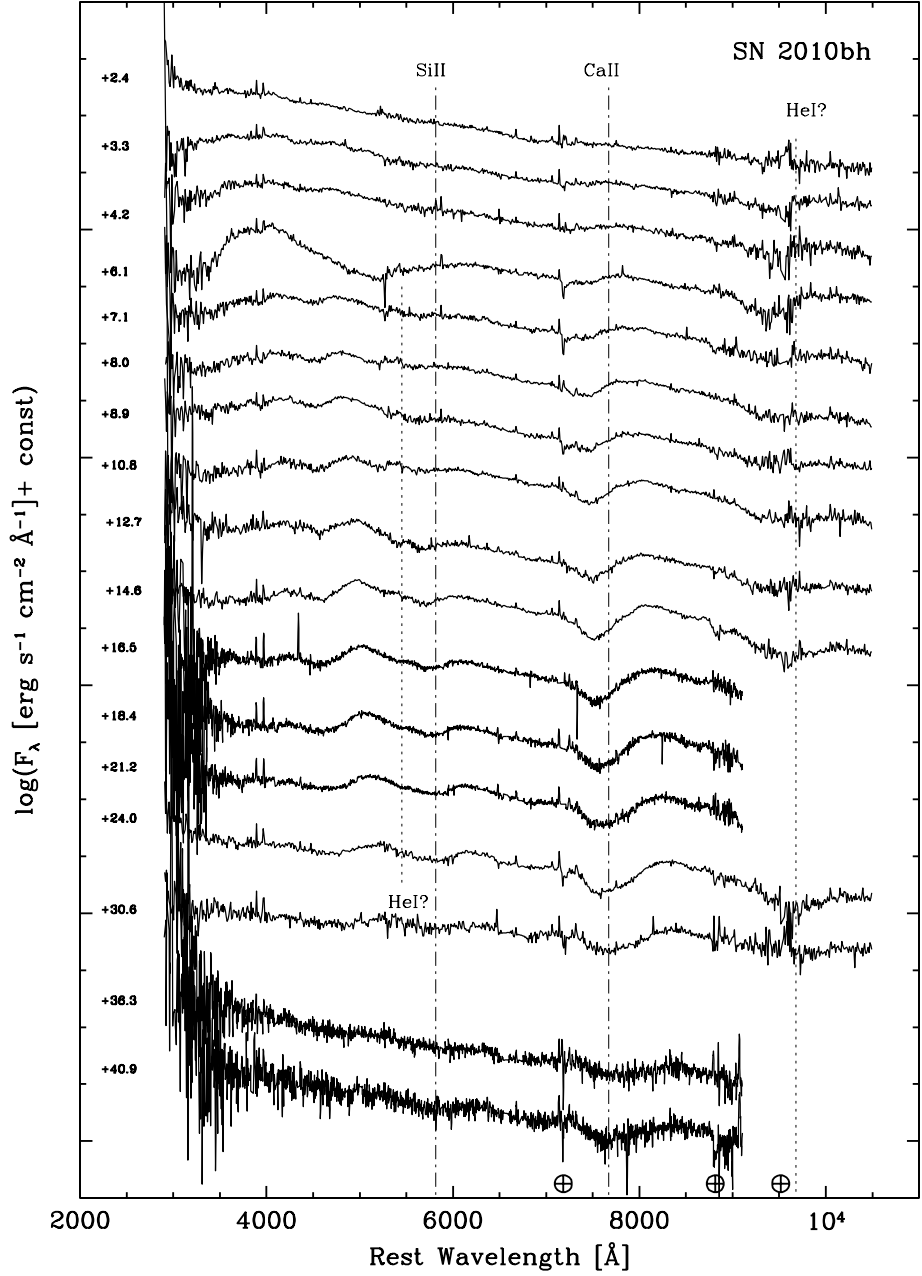}
\caption{Spectral evolution of SN\,2010bh in the optical range. The positions of the tentative identifications of He\,I $\lambda$5876 and He\,I 1.083 $\mu$m  features are marked with dot vertical lines at the wavelength of the minimum on April 11 and 18 spectrum, respectively. Further details are in the caption of Fig. \ref{10bh_spectevol}.
\label{10bh_opt}}
\end{figure}

\begin{figure}
\includegraphics[scale=.7]{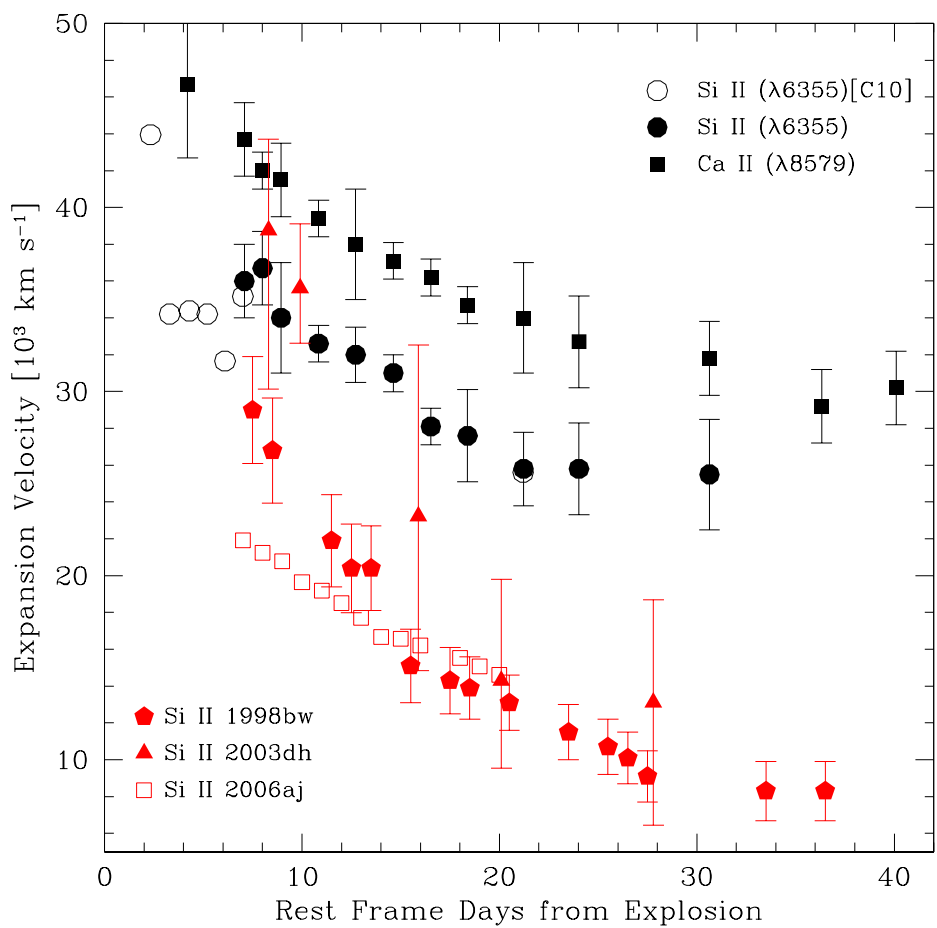}
\caption{Temporal evolution of the expansion velocity of SN\,2010bh measured from different ions. The Si\,II $\lambda$6355 line velocities from \citet{Chornock} are reported with
open circles. The SN\,2010bh Si\,II $\lambda$6355 expansion velocities are compared to those of SNe 1998bw, 2006aj (measurements performed directly on
the spectra published in \citealt{Patat98bw} and \citealt{Mazzali06aj}, respectively) and 2003dh  \citep{Hjorth}. \label{spec_velox}}
\end{figure}

The spectral sequence of SN\,2010bh (Figures  \ref{10bh_spectevol} and \ref{10bh_opt}) is unique among BL-SNe and HNe for its detailed temporal coverage and extended wavelength range (3000--24800 \AA).
At early epochs, the  spectral energy distribution can be fit with a blackbody spectrum with $T_{\rm bb}\approx 8500$\,K.   
Thereafter, the continuum, shaped by broad bumps and absorptions, becomes redder with time because of expansion and cooling.
The two main minima at $\sim 5500$ \AA\ and $\sim 7500$ \AA\  (Fig.\,\ref{10bh_opt}) can be identified with  the blueshifted Si\,II ($\lambda6355$) and Ca\,II NIR triplet  ($gf$-weighted line centroid $\lambda$8579) absorption lines, respectively.
These lines are the most representative of the stratification of the expanding ejecta: their expansion velocities and evolution with time are shown in Figure \ref{spec_velox}.
All velocities have been determined by fitting a Gaussian profile to the absorption features in the rest-frame spectra and measuring the blueshift of the minimum. The uncertainty that affects each line velocity has been taken equal to three times the standard deviation of the measured minimum positions.
The velocity of Si\,II $\lambda$6355 ranges from $\sim$36,000\, km\,s$^{-1}$ at about 7 days from the explosion to $\sim$25,500 km\,s$^{-1}$ at the last epoch. Ca\,II velocities are systematically about 20\% higher than those of Si\,II. In Figure  \ref{spec_velox}, we note that the velocity of the Si\,II $\lambda$6355 line in SN\,2010bh is higher than in SNe 1998bw and 2006aj,  while it is similar to that of SN\,2003dh although with a shallower drop.  

Considering its importance in constraining the nature and the evolutionary state of the progenitor star, we have searched for the spectroscopic signature of helium  in our spectra.   We observe weak  absorption features in the optical and NIR that may be compatible with He\,I $\lambda$5876 and He\,I 1.083 $\mu$m blueshifted by $\sim$20,000--30,000 km\,s$^{-1}$ and $\sim$28,000--38,000 km\,s$^{-1}$, respectively (see Fig. \ref{10bh_opt}). The latter is in agreement with the velocities measured for Si\,II and Ca\,II.  However, He\,I features may be blended with other species, like Na\,I in the optical and C\,I or Si\,I  in the NIR  (\citealt{MazzaliLucy}; \citealt{Millard99}; \citealt{Sauer94I}; \citealt{Taubix04aw}). 
In particular, the contribution of C\,I  cannot be ruled out. Indeed, by comparing the spectra of SN\,2010bh to those of SNe 1998bw \citep{Patat98bw} and 2007gr \citep{Valenti07gr} at similar epochs, we can identify the broad absorption at $\sim 1.6\,\mu$m  with the  C\,I $\lambda$16,890 line.
Such absorption  becomes more prominent starting  8 days after explosion. 
A detection of the He\,I  2.058\,$\mu$m line, typically not blended with other species, in the spectra of SN\,2010bh is not possible, since the line, possibly blueshifted at any velocity up to 35,000\,km\,s$^{-1}$, would lie in the observed range 1.9--1.95\,$\mu$m, which is heavily affected by  telluric absorptions. Consequently, we cannot confirm the identification of  He\,I $\lambda$5876 and He\,I 1.083\,$\mu$m.
Spectral modeling may help in recognizing the different ions contributing to the spectral line formation. This will be the scope of a future paper. 

In \citet{BufanoCBET}, based on the spectrum taken on March 23, we reported the presence of a significant  flux deficit in the range 4500--5500~\AA. The flux density also seems low at wavelengths shorter than $\sim 3500$~\AA. From the spectral evolution (Fig. \ref{10bh_spectevol}), we can see that such deficits are seen only at this epoch. 
A careful analysis of the X-shooter spectrum has not identified any instrumental cause of these features.  
However, their time scale is too short to be explained physically, and therefore we will regard them as spurious.

\begin{figure}
\includegraphics[scale=.7]{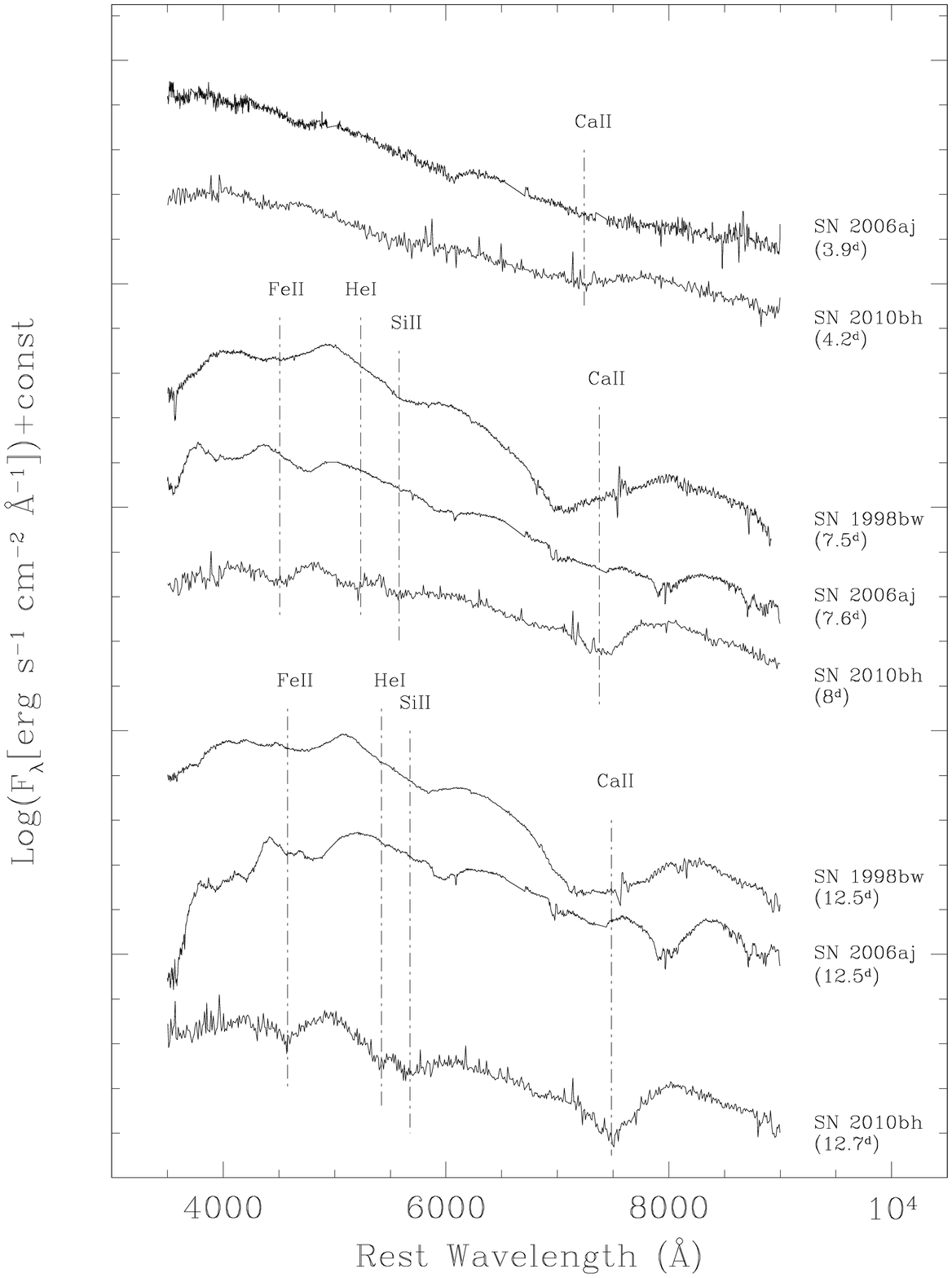}
\caption{  Spectral comparison of SN\,2010bh with other GRB-SNe (\citealt{Patat98bw}; \citealt{Mazzali06aj}). Epochs are reported in rest-frame days after the explosion. \label{spec_comp}}
\end{figure}

In Figure \ref{spec_comp}, we compare the spectra of SN\,2010bh with those of SNe 1998bw \citep{Patat98bw} and 2006aj \citep{Mazzali06aj} at similar phases after explosion.
At  $\sim 4$ days from the burst, both SNe\,2006aj and 2010bh present a featureless spectrum, with the exception in the latter SN of a weak and broad 
($\sim$ 47,000\,km\,s$^{-1}$) P-Cygni feature due to the Ca\,II NIR triplet.
This line becomes more prominent at later phases and displays a decreasing velocity (see also Fig. \ref{spec_velox}). The Ca\,II triplet velocity remains   significantly higher than in SN\,2006aj, for which \citet{Mazzali06aj} measured $\sim$25,000\,km\,s$^{-1}$ roughly constant with time.  In SN\,1998bw spectra,  \citet{Patat98bw} found that the main contribution to the absorptions at $\sim$ 7000 \AA\  was given by O\,I (Fig.\ref{spec_comp}), which is not obvious in our spectra of SN\,2010bh.
Absorptions at $\sim 3500$\,\AA\ and $\sim 4500$\,\AA\ in SN\,2010bh spectra are likely due to Fe\,II and Ti\,II, as found for SN\,2006aj through spectral modeling \citep{Mazzali06aj}.

\subsection{Spectrum in the Nebular Phase} \label{neb_spec}

\begin{figure}
\includegraphics[scale=.7, angle=90]{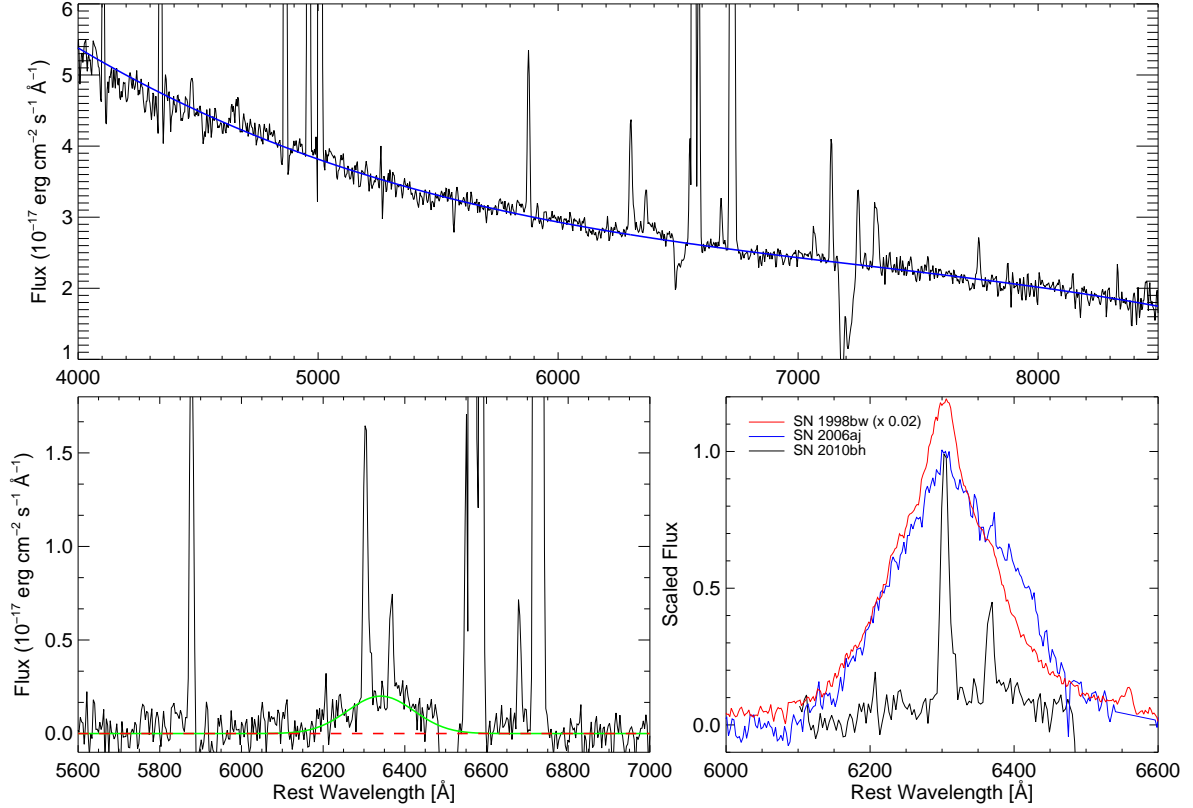}
\caption{{\it (Upper Panel)}  FORS2 spectrum of SN\,2010bh   at  $\sim$186 rest-frame days after the explosion obtained by combining the  spectra acquired in September and October 2010 (see Table \ref{SN2010bh_spec}). The blue line is the fit of the continuum we used to subtract the host-galaxy continuum flux contamination.  
{\it (Lower-left panel)}  Close-up view of the continuum-subtracted nebular spectrum in the wavelength interval 5500--7000\,\AA, centered on the expected wavelength of the [O\,I] $\lambda\lambda$6300, 6363 nebular emission lines. Lines from the host galaxy H\,II region dominate. The broad emission at $\sim 6300$\,\AA\ is fitted with a Gaussian curve (green line). {\it (Lower-right panel)} The spectra of SN\,1998bw at $\sim 214$ rest-frame days after explosion (red line; \citealt{Patat98bw}) and SN\,2006aj at 206 days (blue line; \citealt{Mazzali06ajlate})  are shown for comparison. The fluxes of SNe\,1998bw  and 2010bh have been normalized to that of SN\,2006aj at the peak of the [O\,I] line (the narrow H\,II region component, in the case of SN\,2010bh). An additional rescaling of SN\,1998bw has been done for clarity.   \label{late_spec}}
\end{figure}

In the upper panel of Figure \ref{late_spec}, we show the FORS2 nebular spectrum ($\sim$186 days after explosion in the rest frame).  
At this time, no significant continuum flux contribution from the supernova photosphere is expected. 
Therefore, since the SN exploded on a bright region of the host galaxy, we fit the spectral continuum with a polynomial function to obtain and subtract the background flux.  
The final continuum-subtracted nebular spectrum in the  range 5500--7000\,\AA\ is plotted in the lower-left panel of Figure \ref{late_spec}.   
It shows the [O\,I]  narrow emission lines  at 6300\,\AA\  and 6363\,\AA\  from the underlying galaxy region,  and a broad but faint component, which, when fitted with a Gaussian function, peaks at 6340\,\AA\ with a total flux of $1.3\times10^{-16}$ erg cm$^{-2}$ s$^{-1}$.

 From the comparison of SN\,2010bh with SN\,1998bw and SN\,2006aj at similar rest-frame phases after the burst (214 and 206 days, respectively; \citealt{Patat98bw}; \citealt{Mazzali06ajlate}; lower-right panel in Fig. \ref{late_spec}), we find  that the  [O\,I]  bump is very weak in SN\,2010bh. The signal in the continuum-subtracted spectrum is too low to guarantee a secure measurement of the [O\,I] abundance and to perform spectral modeling.
However, taking into account the lack of strong evidence of O\,I lines in SN\,2010bh spectra at early epochs, this could indicate  a very small amount of ejected oxygen, and, consequently, a less massive progenitor than for SNe\,1998bw and 2006aj.\
On the other hand, it could also be explained as a lower nebular flux,  providing an indirect additional support for the   $M_{\rm Ni}$, $M_{\rm ej}$, and  $E_{\rm k}$ values we found and discuss in the next Section (see also Table \ref{par}).  Indeed, considering that the energy input scales roughly as $M_{\rm Ni} \times (\tau_{\gamma} + 0.035)$, where $\tau_{\gamma}$ is the optical depth to radioactive gamma rays ($\tau_{\gamma} \approx 10^3 \, M_{\rm ej}^2\, E_{\rm k}^{-1}\, t^{-2}$: see, e.g., \citealt{Maeda2003}) and 0.035 is the positron contribution,  we obtain a late-time flux in SN\,2010bh which is about a factor of $\sim$7 and a factor of $\sim$2.5 smaller than that in SN\,1998bw and SN\,2006aj, respectively (for the same distance and reddening). 

\subsection{Bolometric Light Curve}

\begin{figure}
\includegraphics[scale=.7]{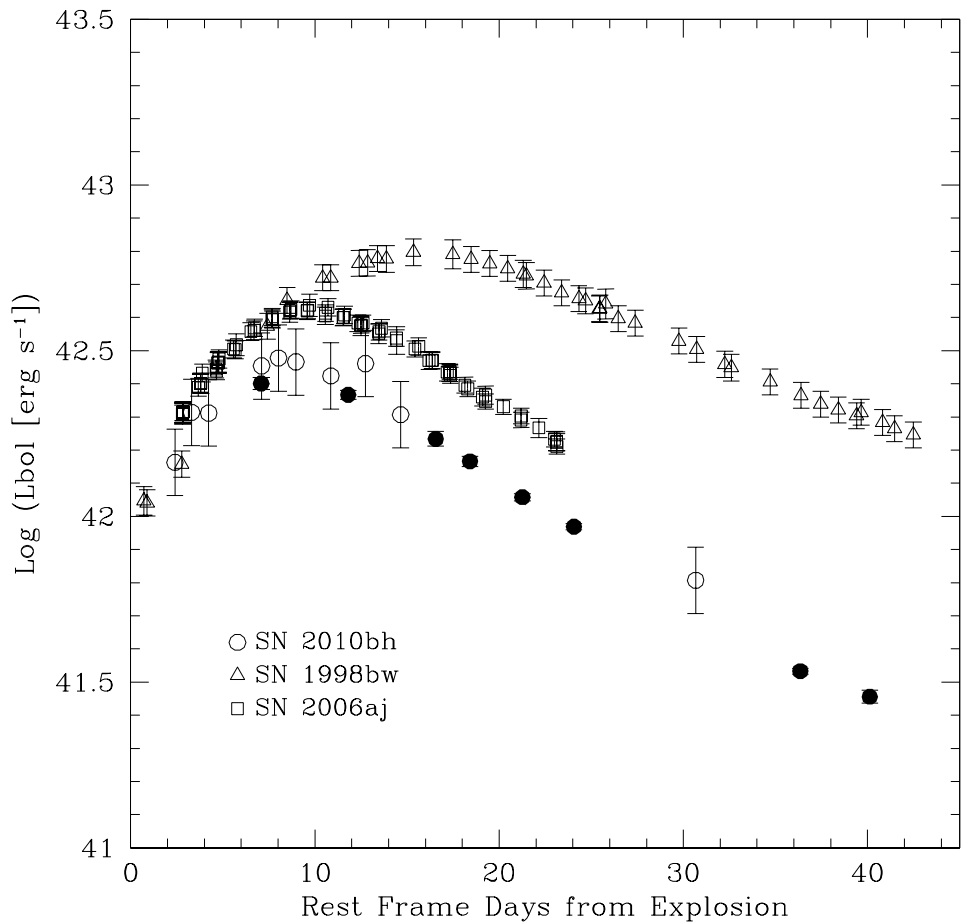}
\caption{SN\,2010bh pseudo-bolometric ($BVRI$) light curve compared with those of SNe 1998bw and 2006aj (open triangles and squares, respectively). SN\,2010bh pseudo-bolometric luminosity obtained from FORS2 photometry is shown with filled circles, while open circles represent synthetic magnitudes obtained from X-shooter spectra.  \label{label_bol}}
\end{figure}

Since photometry in the individual bands is affected by possible spectral lines and their time evolution, it is important to construct a bolometric light curve to estimate reliably the SN physical parameters.
As a first approximation, we obtained a quasi-bolometric BVRI light curve using FORS2 photometry and the synthetic BVRI magnitudes obtained from X-shooter spectra.
To this aim, BVRI magnitudes were firstly corrected for extinction and  converted to flux density at the effective wavelength of the
Johnson-Cousin filters. The spectral energy distribution was then  integrated over the entire wavelength range and, finally, the 
integrated flux was converted into luminosity using the adopted distance (Sect. \ref{host}). The error bars of the bolometric luminosities obtained using X-shooter spectra are equal to the typical uncertainty of 10\% that affects spectra flux calibration.\\
In Figure \ref{label_bol}, the pseudo-bolometric luminosity of SN\,2010bh is compared to those of the GRB-SNe 1998bw and 2006aj.  It
displays an evolution similar to that of SN\,2006aj, although with a fainter peak by about 0.2 dex ($L_{\rm bol, 10bh} \approx 3 \times 10^{42}$ erg s$^{-1}$).\\
For the  bolometric light curve fitting, we used a a simple model, that assumes, for the photometric phases, a concentration of the radioactive nickel ($^{56}$Ni) in the core, a homologous expansion of the ejecta and a spherical symmetry, following the prescriptions of \citet{Arnett82}, and, for the nebular phases,  includes the energy contribution from the $^{56}$Ni$-^{56}$Co$-^{56}$Fe decay (\citealt{Sutherland}; \citealt{Cappellaro}).
 For a detailed description of the model see  \citet{Valenti03jd}.
The bolometric light curve model suggests a total ejected mass of radioactive $^{56}$Ni  of $M_{\rm Ni} = 0.12 \pm 0.02\,{\rm M}_\odot$.  
The lack of measurements during the  nebular phase prevents us from verifying this value through the radioactive tail.  \\
Although a direct comparison with the quasi-bolometric curves created by \citet{Cano} and \citet{Olivares} is not possible because of the difference in the wavelength ranges used  to construct them, our $M_{\rm Ni}$ estimate is in good agreement with that of \citet{Cano} who found a $^{56}$Ni mass of $M_{\rm Ni} = 0.10 \pm 0.01\,{\rm M}_\odot$.  \citet{Olivares} derived a higher value  ($M_{\rm Ni} = 0.21 \pm 0.03\,{\rm M}_\odot$), likely because of the higher extinction correction.

\section{Discussion and Conclusions}\label{discussion}  
In order to obtain further information on the explosion and stellar progenitor, we compared the observed properties of SN\,2010bh with those of SN\,2006aj. 
We find that the SN\,2010bh $R$-band light-curve width was $\tau_{\rm peak, 10bh} = (0.96 \pm 0.11) \times \tau_{\rm peak, 06aj}$ 
by ''stretching'' its time-scale in order to match SN\,2006aj light curve (see \citealt{Perlmutter}) and its photospheric expansion velocity  $v_{\rm ph, 10bh} = (1.74 \pm 0.05) \times v_{\rm ph, 06aj}$ (assuming Si\,II $\lambda$6355 as a good tracer, see e.g. \citealt{Valenti03jd}).
Using the relations between $\tau_{\rm peak}$ and $v_{\rm ph}$ with the ejected mass $M_{\rm ej}$ and the kinetic explosion energy $E_{\rm k}$ ($\tau_{\rm peak} \varpropto M_{\rm ej}^{3/4}E_{\rm k}^{-1/4}$; $v_{\rm ph} \varpropto M_{\rm ej}^{-1/2}E_{\rm k}^{1/2}$; Arnett 1982, 1996) and the $M_{\rm ej}$ and $E_{\rm k}$ estimates of SN\,2006aj found by \citet{Mazzali06aj}, we derive $M_{\rm ej} \approx (3.2 \pm 1.6)\, {\rm M}_\odot$ and  $E_{\rm k} \approx (9.7 \pm 5.5)\times 10^{51}$\, erg. 
From the analytical modeling of their pseudo-bolometric light curves, \citet{Cano} and \citet{Olivares} found comparable ejected masses:  
$M_{\rm ej} = 2.24 \pm 0.08\, {\rm M}_\odot$ and $M_{\rm ej} = 2.6 \pm 0.2\, {\rm M}_\odot$, respectively.
While the kinetic energy found by Cano et al. is not too dissimilar from ours ($E_{\rm k} = (1.39 \pm 0.06) \times 10^{52}$\, erg), the one 
derived by Olivares et al. is significantly larger, $E_{\rm k} = (2.4 \pm 0.7) \times 10^{52}$\, erg.  These discrepancies are related to the different species used for the velocity measurements and to the uncertainties of the measurements themselves, that are affected by line blending and by some arbitrariness in the choice of the line profiles.  These problems are overcome by the use of a radiative transport model.

\begin{figure}
\includegraphics[scale=.7]{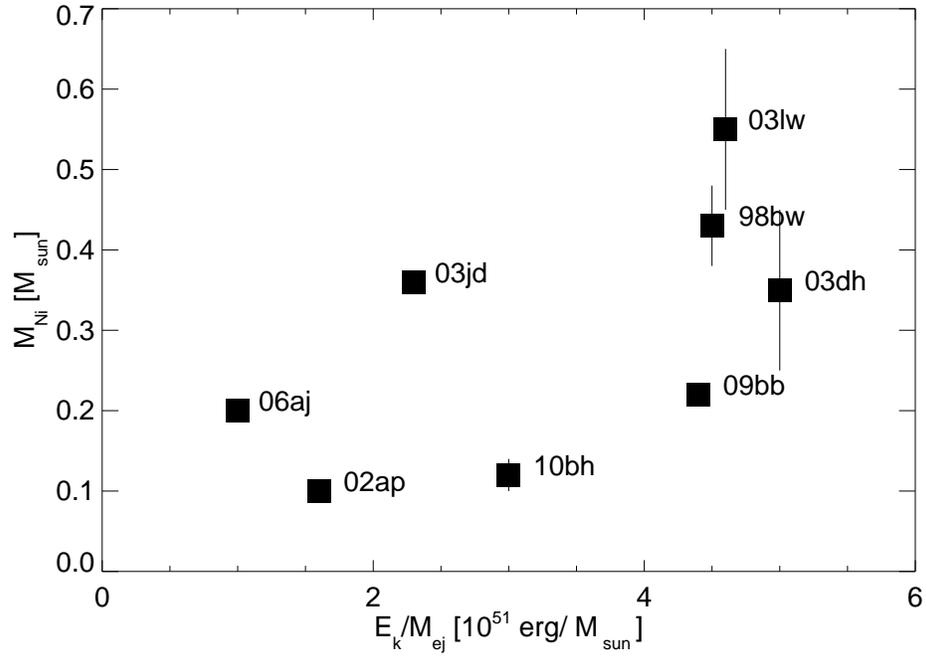}
\caption{The ejected  $^{56}$Ni mass as a function of the ratio between the  explosion energy and the ejected mass ($E_{\rm k}/M_{\rm ej}$) for several broad-lined supernovae/hypernovae. \label{energetic}}
\end{figure}

\begin{figure}
 \begin{center}
 \includegraphics{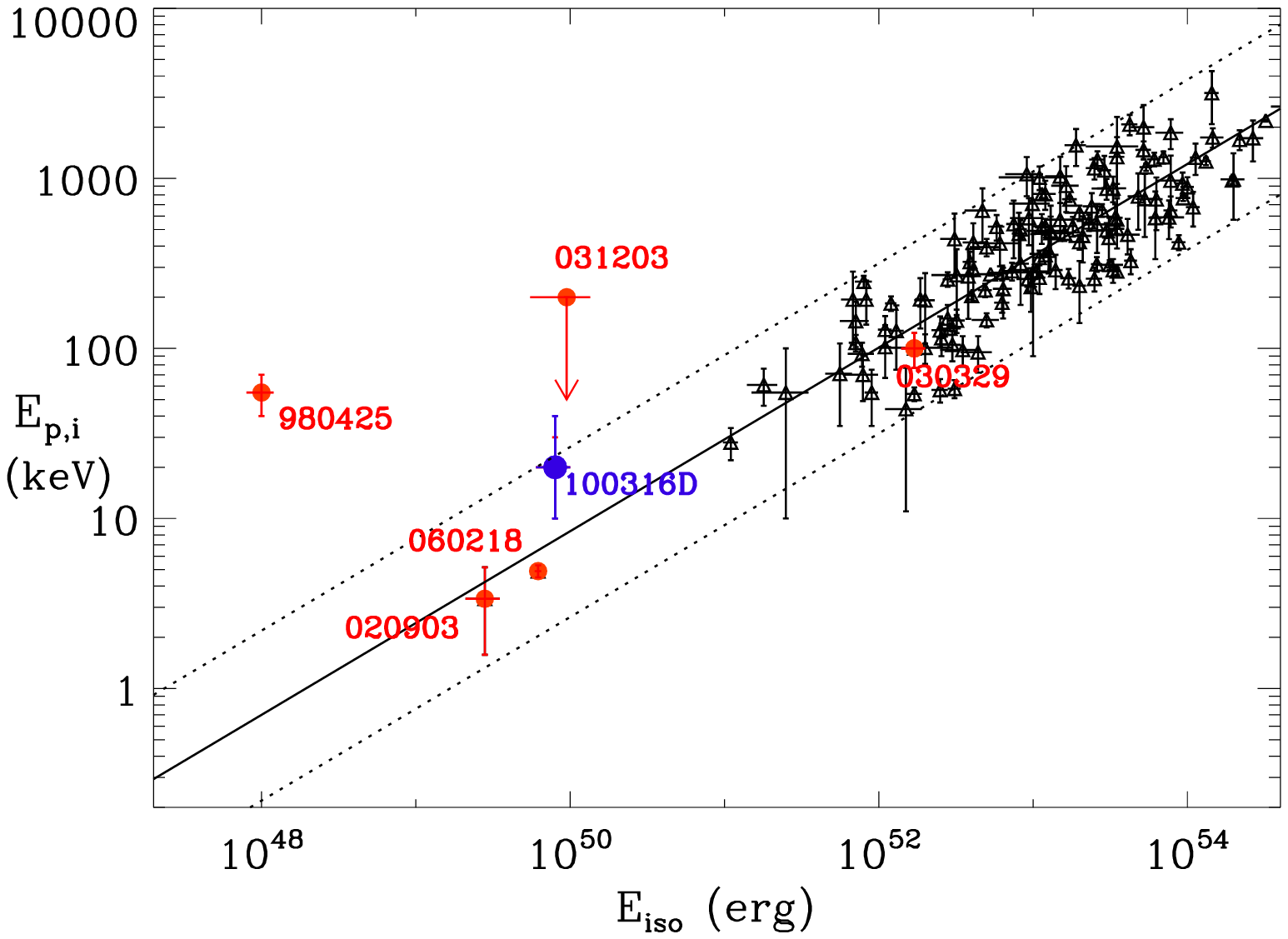}
 \caption{ Location of GRB\,100316D in the $E_{\rm p,i}-E_{\rm iso}$ plane. GRBs/XRFs connected with a spectroscopically confirmed SN are shown with red dots. Similar to GRB\,060218, GRB\,100316D is consistent with the correlation  $E_{\rm p,i}-E_{\rm iso}$ (solid line) derived  by \citet{Amati02}. The two parallel dotted lines delimit the 2.5$\sigma$ confidence region \citep{Amati09}.  \label{Amati}   }
 \end{center}
 \end{figure}

Although the photometric evolution of SN\,2010bh was similar to that of SN\,2006aj (i.e., similar light-curve width), SN\,2010bh had a higher $M_{\rm ej}$ and $E_{\rm k}$, explaining the faster expansion velocities measured from the spectra. This could suggest that spectra are more sensitive than light curves to possible effects of the viewing angle, in case of asymmetric explosion (a bipolar explosion is expected in the presence of a GRB; \citealt{PiranGRB}).
Then we would expect that the weakest GRB (most off-axis) also has the lowest registered $E_{\rm k}/M_{\rm ej}$ ratio.
In Table \ref{par}, we report the $M_{\rm ej}$ and $E_{\rm k}$ values found for previous GRB-SNe and other broad-lined SNe~Ic, as well as the corresponding $E_{\rm k}/M_{\rm ej}$ ratio, and thus plot the latter versus the relative ejected $M_{\rm Ni}$ (Fig. \ref{energetic}).
While SN\,2010bh has an intermediate $E_{\rm k}/M_{\rm ej}$ ratio, it lies on the low Nickel mass tail of the energetic type Ic SN distribution (SN\,2002ap, \citealt{Mazzali02ap}; SN\,2003jd, \citealt{Valenti03jd}). 
The main reasons for such a wide variety among GRB-SNe cannot rely only on differences in the viewing angle, but must be intrinsic (e.g., explosion collimation, progenitor mass, etc.).
Indeed, GRB-SNe have been supposed to come from different explosion scenarios (see \citealt{Woosley}, and references therein), where the core collapse of a massive progenitor star (20--60\,M$_\odot$) leads to the formation of different central engines (a magnetar  or a black hole).
On the other hand, this heterogeneity does not have an obvious correspondence in the properties of the GRB.  In Figure \ref{Amati} the intrinsic peak energy is plotted  as a function of the isotropic emitted energy (see, e.g., \citealt{Amati09}), showing  that GRB-SNe events, including GRB\,100316D,  are consistent  with the correlation  in the $E_{\rm p,i}-E_{\rm iso}$ plane holding for all long GRBs (with exception of GRB\,980425).  As noted by \citet{Starling}, GRB\,100316D has prompt gamma-ray spectral properties similar to GRB\,060218.

The similarity between the light curves of SN\,2010bh and  SN\,2006aj and between the observational characteristics of their associated GRBs suggests  a common explosion mechanism for these two events. 
SN\,2010bh would be produced by the core collapse of a relatively massive progenitor star (20--25\,M$_\odot$), which leaves behind a magnetar, similar to SN\,2006aj \citep{Mazzali06aj}.
Indeed, by analyzing the spectral and temporal properties of the associated GRB\,100316D, \citet{Fan}  claimed that SN\,2010bh was possibly powered by a magnetar with a spin period of $P \approx 10$ ms and a magnetic field $B \approx 3 \times 10^{15}$\,G.
In this picture, the magnetar rotational energy would be injected in the expanding remnant on a timescale set by the magnetic dipole radiation.
Following Eq. 2 of \citet{Kasen}, we would expect $t_0 \approx 12$ hr; thus, even if a good fraction of the pulsar energy is converted into radiation, it will be lost by adiabatic expansion before escaping from the ejecta.
\citet{Maeda2007} argued that in SN\,2006aj the contribution from the magnetar to the light curve was likely negligible compared with the $^{56}$Ni power, and this could be applicable in general to all magnetar powered GRBs/XRFs. 

We consider the detection at $\sim 12$\,hr after the burst ($R \approx 21.4$ mag) as the contribution from the ``cooling envelope" during the post- shock breakout phases (we may also have a nonthermal contribution from the afterglow at this phase).  Following \citet{Chevalier},  we estimate the progenitor radius: using their Eq. 4 and 5 and $E_{\rm k}$ and $M_{\rm ej}$ from our Table \ref{par}, we obtain the expected blackbody luminosity and temperature at 12\,hr  for different progenitor radii. Then the expected blackbody spectrum was converted to the observed frame (assuming the SN\,2010bh redshift and total color excess) and convolved with the $R$-band filter.
The resulting magnitude was fainter than the observed one only in the case of a progenitor radius $R \lesssim 10^{11}$ cm.
Even considering the 3-sigma upper limit flux, we would obtain a progenitor radius of the same order.
This estimate is certainly  rough, since it is based only on the single $R$-band image and uses simplified formulae \citep{Chevalier}, but  it could provide support for a compact progenitor scenario and it is also consistent with  the  initial radius of $7\,\times\,10^{11}$\,cm found by  \citet{Olivares} analyzing the early X-ray-to-NIR emission. 

On the other hand, we cannot exclude a scenario in which the outcome of GRB\,100316D/ SN\,2010bh was a black  hole, which could explain the small ejected mass as a consequence of the possible fall-back of ejecta onto the BH. Recently, such small ejected masses have been inferred from the kinematics of black holes with $ > $10 M$_\odot$ in our Galaxy, such as Cygnus X-1 (\citealt{Mirabelx1}; \citealt{Goux1}), GRS\,1915 and V404\,Cyg \citep{MirabelBH}.
However, in this case it may be  difficult to accommodate the higher expansion velocities measured for  SN\,2010bh than those of previous HNe with a collapsar progenitor (e.g., SN\,1998bw).

As anticipated, the high $v_{\rm ph}$ may likely be explained as an effect of the explosion geometry. Good indicators of the explosion geometry are the nebular emission lines of Fe\,II (a blend near 5200\,\AA) and [O\,I] ($\lambda\lambda$6300,6363;  \citealt{Maeda2002}, 2006; \citealt{MazzaliNeb}), whose profiles can reveal the presence of asymmetry.  No such information on the SN\,2010bh explosion geometry can be deduced from its nebular spectrum, because of  the faintness of the emission lines.

\acknowledgments
We thank  the ESO shift coordinators and night assistants for their excellent support during the observational campaign:
 A. Alvarez, T. Rivinius, T. Szeifert, T. Bensby, C. Ledoux, 
J. Smoker, C. Melo, F. J. Selman, S. Mieske, S. Stefl, P. Lynam, and V. D. Ivanov.
We thank L. Koopmans for allowing us to activate our program during his scheduled observing time on 2010 March 18.
F.B. is grateful to S. Taubenberger for fruitful discussion. F.B., S.B., E.C., S.V., and M.Turatto are supported by grant ASI-INAF I/009/10/0.
Financial support from PRIN INAF 2009 is acknowledged.  E. Pian acknowledges
hospitality at the ESO HQs in Santiago, where part of this work was accomplished.
G.P. acknowledges support by the Proyecto FONDECYT 11090421, Comit\'e Mixto ESO-Gobierno de Chile, Millennium Center for
Supernova Science through grant P06-045-F funded by ÒPrograma Iniciativa Cient\'ifica Milenio de MIDEPLAN.Ó, Centro de Astrof«õsica FONDAP 15010003, Center of Excellence in Astrophysics and Associated Technologies PFB 06 and Proyecto interno UNAB N¡ DI-28-11/R.
R.L.C.S. is supported by a Royal Society Fellowship. A.V.F.'s work has been funded by NSF grant AST--0908886 and NASA/{\it Swift} grant NNX10AI21G
M.Tanaka, K.M., and K.N. are grateful to the World Premier International Research Center Initiative, MEXT, Japan.
The Dark Cosmology Centre is funded by the DNRF. JPUF acknowledges support form the ERC-StG grant EGGS-278202.

{\it Facilities:} \facility{VLT/X-shooter, VLT/FORS2, PROMPTs}.

\appendix

\begin{table}
\begin{center}
\caption{Journal of late-epoch template observations.\label{templates}}
\begin{tabular}{ccccc}
\hline\hline
UT & JD & Phase$^{\dagger}$ Instr. & Bands\\
{}& (+2400000)&&&\\
\hline
2010/09/17 &55456.8& 184.8 & FORS2 & $UBVRI$\\
2011/01/26 &55587.5 & 315.5 &PROMPT5 & $I$ \\
2011/02/02 &55594.5 & 322.5&PROMPT1 & $VRI$\\
2011/02/02 &55594.5& 322.5 &PROMPT5 & $VR$\\
\hline
\end{tabular}
\begin{minipage}{13cm}
$^{\dagger}$ {\footnotesize Phases from {\it Swift}/BAT trigger (2010 March 16.53; \citealt{BATtrigger}).}
\end{minipage}
\end{center}
\end{table}

\begin{table}
\begin{center}
\caption{Optical magnitudes of the local reference stars in the field of SN\,2010bh. \label{SeqStars}}
\begin{tabular}{cccccc}
\tableline\tableline
Star ID & $U$ & $B$ & $V$ & $R$ & $I$ \\
\tableline
 1& 22.79$\pm$0.06& 21.74$\pm$0.02&20.31$\pm$0.03& 19.43$\pm$0.01& 18.64$\pm$0.02\\
 2&19.88$\pm$0.04& 19.42$\pm$0.02&18.57$\pm$0.04& 18.02$\pm$0.01& 17.51$\pm$0.01\\
 3&21.48$\pm$0.09& 21.57$\pm$0.07&20.82$\pm$0.04& 20.35$\pm$0.05& 19.86$\pm$0.03\\
 4&18.84$\pm$0.06& 18.35$\pm$0.02&17.55$\pm$0.03& 17.05$\pm$0.01&16.57$\pm$0.01\\
 5&21.70$\pm$0.10& 21.86$\pm$0.02&21.33$\pm$0.03&21.33$\pm$0.03& 20.50$\pm$0.02\\
 6&21.24$\pm$0.12& 20.78$\pm$0.02&19.98$\pm$0.03&19.45$\pm$0.01& 18.98$\pm$0.01\\
 7&17.49$\pm$0.07& 17.49$\pm$0.03&16.96$\pm$0.02&16.60$\pm$0.06& 16.23$\pm$0.07\\
 8&$\cdots$&                         22.58$\pm$0.08&20.98$\pm$0.03&20.02$\pm$0.01& 19.04$\pm$0.02\\
 9&18.30$\pm$0.06& 17.95$\pm$0.02&17.18$\pm$0.03&16.72$\pm$0.03& 16.25$\pm$0.03\\
 10&$\cdots$                      &21.23$\pm$0.08&19.78$\pm$0.04&18.88$\pm$0.01& 18.14$\pm$0.01\\
\tableline
\end{tabular}
\end{center}
 {\footnotesize No corrections have been applied to the reported magnitudes.}
\end{table}

\begin{deluxetable}{ccccccccc}
\tabletypesize{\scriptsize}
\tablecaption{$UBVRI$  observed magnitudes of SN\,2010bh.  \label{SN2010bh_phot}}
\tablehead{\colhead{UT} & \colhead{JD} & \colhead{Phase$^{\dagger}$} &\colhead{$U$}&\colhead{$B$} &\colhead{$V$}&\colhead{$R$} &\colhead{$I$}&\colhead{Instr. }\\
& \colhead{+2400000}& && && &&}
\startdata
2010/03/17 &  55272.5&0.4 &  -- -- & -- -- & -- --&	${21.49}_{ +0.23}^{ -0.37}$ & -- -- &X-shooter\\
2010/03/19 &  55274.5&2.4 &  -- -- & -- -- & -- --&	$20.24_{+0.20}^{-0.24}$  & -- -- &X-shooter\\
2010/03/20 &  55275.5&3.3 &  -- -- & -- -- & -- --&	$19.91_{+0.11}^{-0.14}$  & -- -- &X-shooter\\
2010/03/21&   55276.5&4.2 &  -- -- & -- -- & -- --&	$19.88_{+0.16}^{-0.14}$  & -- -- &X-shooter\\
2010/03/24 &  55279.5&7.1 &  -- -- & -- -- &  $20.01_{+0.10}^{-0.11}$& $19.63_{+0.04}^{-0.06}$&$19.54_{+0.31}^{-0.33}$&FORS2 \\
2010/03/24 &  55279.5&7.1 &  -- -- & -- -- & -- --&        $19.48_{+0.51}^{-0.53}$  & -- -- &X-shooter\\
2010/03/25 &  55280.5&8.0 &  -- -- & -- -- & -- --&	$19.37_{+0.06}^{-0.08}$  & -- -- &X-shooter\\
2010/03/26 &  55281.5&8.9 &  -- -- & -- -- & -- --&	$19.47_{+0.10}^{-0.12}$  & -- -- &X-shooter\\
2010/03/26&   55281.7&9.1 &  -- -- & -- -- &   19.66 $\pm$ 0.21& 19.21 $\pm$ 0.14 & -- -- &PROMPT5$^*$ \\
2010/03/28 &  55283.5&10.8&  -- -- & -- -- & -- --&	$19.47_{+0.06}^{-0.08}$  & -- -- &X-shooter\\
2010/03/28 & 55283.5 &10.8&  -- -- & -- -- & -- -- &  19.40 $\pm$ 0.15 & 18.93$\pm$ 0.12 &PROMPT5$^*$\\
2010/03/28 & 55283.5 &10.8&  -- -- & -- -- &    19.73 $\pm$ 0.15 &  19.36 $\pm$ 0.11 &   19.00 $\pm$ 0.16  &PROMPT1$^*$\\
2010/03/29 & 55284.5 &11.8&  -- -- & 21.05 $\pm$ 0.08 & $20.05_{+0.10}^{-0.11}$   & $19.65_{+0.08}^{-0.10} $&  $19.25_{+0.11}^{-0.12}$ &FORS2 \\
2010/03/29 & 55284.5 &11.8&  -- -- & -- -- &  19.99 $\pm$ 0.15 &  19.41 $\pm$ 0.11 & -- -- &PROMPT1$^*$ \\
2010/03/30 & 55285.5 &12.7 & -- -- & -- -- & -- --&      -- --     & $19.09_{+0.167}^{-0.180} $&X-shooter\\  
2010/03/30 & 55285.5 &12.7&  -- -- & -- -- & -- -- &19.50 $\pm$ 0.15 &  19.05 $\pm$ 0.12 &PROMPT5$^*$ \\
2010/03/31 & 55286.5 &13.7&  -- -- & -- -- &  20.17 $\pm$ 0.19  & 19.49 $\pm$ 0.12  & 19.10 $\pm$ 0.14&PROMPT1$^*$ \\
2010/04/01 & 55287.5 &14.6&  -- -- & -- -- & -- --&	$19.77_{+0.07}^{-0.07}$  & -- -- &X-shooter\\ 
2010/04/03 & 55289.5 &16.5&  20.85 $\pm$ 0.37 & 21.62 $\pm$ 0.06    &$20.38_{+0.16}^{-0.18}$&$19.86_{+0.22}^{-0.25}$ &$19.39_{+0.08}^{-0.09}$& FORS2 \\
2010/04/05  &55291.5 &18.4&  21.60 $\pm$  0.21 &21.99 $\pm$ 0.11    &$20.50_{+0.13}^{-0.14}$ & $19.96_{+0.13}^{-0.15}$ &$19.55_{+0.09}^{-0.11}$ & FORS2\\
2010/04/05 & 55291.5 &18.4&  -- -- & -- -- &20.49 $\pm$ 0.23 & -- -- & -- --  &PROMPT5$^*$ \\
2010/04/08  &55294.5 &21.2&  22.14 $\pm$  0.60&22.58 $\pm$ 0.07    &$20.83_{+0.12}^{-0.15}$ & $20.18_{+0.06}^{-0.09} $ & $19.60_{+0.02}^{-0.04}$ & FORS2\\
2010/04/11  &55297.5& 24.0&  22.34 $\pm$  0.51 &22.64 $\pm$ 0.10   & $21.19_{+0.07}^{-0.10}$  &$ 20.38_{+0.05}^{-0.09}$ & $19.92_{+0.07}^{-0.10}$ & FORS2 \\
2010/04/18 & 55304.5 &30.6&  -- -- & -- -- & -- --&                                                                                      	 $21.02_{+0.17}^{-0.23} $ & -- -- &X-shooter\\ 
2010/04/24  &55310.5 &36.3&  -- -- &                               23.40 $\pm$ 0.07    &$22.02_{+0.04}^{-0.11}$  &$ 21.44_{+0.06}^{-0.17}$ &$ 21.11_{+0.07}^{-0.15} $   & FORS2 \\
2010/04/28  &55314.5 &40.1&  -- -- & 22.8$^{\ddagger}$ &                               $22.48_{+0.09}^{-0.19}$ & $ 21.66_{+0.11}^{-0.23}$  &$ 21.04_{+0.18}^{-0.27}$ & FORS2  \\
2010/05/18  &55334.5& 59.0 & -- -- & 23.4$^{\ddagger}$&                              $22.89_{+0.13}^{-0.28}$  &$ 22.13_{+0.54}^{-0.72}$&  $21.90_{+0.29}^{-0.47}$ & FORS2   \\
2010/09/17 &55456.8&166.5&22.7$^{\ddagger}$&21.7$^{\ddagger}$&21.9$^{\ddagger}$&21.4$^{\ddagger}$&21.3$^{\ddagger}$& FORS2\\
\enddata
\tablecomments{$^{\dagger}$ Phases from {\it Swift}/BAT trigger (2010 March 16.53; \citealt{BATtrigger}) in the host-galaxy rest frame.  No corrections have been applied to the reported magnitudes, with exception of the ($^{*}$) S-correction for the PROMPT ones.  $^{\ddagger}$Upper limit.  }
\end{deluxetable}

\begin{deluxetable}{cccccc}
\tablewidth{0pt}
\tablecaption{Journal of spectroscopic observations of SN\,2010bh.\label{SN2010bh_spec}}
\tablehead{
\colhead{UT}  & \colhead{Phase\tablenotemark{\dagger} [days]}      &\colhead{Tel./Instr.}  & \colhead{Arm/Grism\tablenotemark{\ddagger}}        & \colhead{Exptime [s]}    & \colhead{Airmass}}
\startdata
2010/03/19.0   & 2.4 & VLT/X-shooter& UV/VIS/NIR &  2400/2400/2400&1.18 \\
2010/03/20.0   & 3.3 & VLT/X-shooter& UV/VIS/NIR &      2600/2600/3240&1.17 \\
2010/03/21.0    & 4.2& VLT/X-shooter& UV/VIS/NIR &      2600/2600/3240&  1.17  \\					                     
2010/03/23.0       &6.1 & VLT/X-shooter & UV/VIS/NIR &     2600/2600/3240& 1.20 \\
2010/03/24.0       & 7.1& VLT/X-shooter& UV/VIS/NIR &    2600/2600/3240&  1.20 \\
2010/03/25.0       &8.0& VLT/X-shooter&UV/VIS/NIR&     2400/2400/2400& 1.18 \\
2010/03/26.0    & 8.9& VLT/X-shooter & UV/VIS/NIR &      2600/2600/3240&    1.18\\
2010/03/28.0     & 10.8& VLT/X-shooter & UV/VIS/NIR &     2400/2400/2400&  1.19\\
 2010/03/30.0	    & 12.7& VLT/X-shooter& UV/VIS/NIR &      2600/2600/3240&   1.19\\
 2010/04/01.0	   & 14.6& VLT/X-shooter& UV/VIS/NIR &      2400/2400/2400& 1.21\\
2010/04/03.0    &16.5& VLT/FORS2& 300V&  1800 &       1.28 	\\				                     
2010/04/05.0     &18.4& VLT/FORS2& 300V&   1800 &         1.22	\\
2010/04/08.0    &21.2& VLT/FORS2& 300V&  1800 &         1.26 	\\
2010/04/11.0     & 24.0& VLT/X-shooter&UV/VIS/NIR &  2400/2400/2400& 1.20	 \\
2010/04/18.0     & 30.6& VLT/X-shooter& UV/VIS/NIR  &  2600/2600/3240&     1.22 \\
2010/04/24.0  &36.3& VLT/FORS2& 300V&  1800 &        1.34\\					                     
2010/04/28.0  &40.9& VLT/FORS2& 300V&  3600 &  1.33 	\\
2010/09/28.6 &195.8& VLT/FORS2& 300V&   3600 &1.60\\	
2010/09/29.6 &196.7& VLT/FORS2& 300V&   7200 &1.80\\	
2010/10/01.7 &198.8& VLT/FORS2& 300V&  1260 &1.47\\		
\enddata 
\tablenotetext{\dagger}{Phases from {\it Swift}/BAT trigger (2010 March 16.53; \citealt{BATtrigger}) in the host-galaxy rest frame. }
\tablenotetext{\ddagger}{X-shooter arm wavelength ranges are UV 3000--5600\,\AA, VIS 5500--10200\,\AA, and NIR 10200--24800\,\AA. FORS2 Grism 300V is 3300--9000\,\AA.}
\end{deluxetable}

\begin{table*}
\begin{center}
\caption{Main physical parameters for GRB-associated SNe and broad-lined SNe Ic.}
\label{par}
\begin{tabular}{cccccc}
\hline\hline
& $E_{\rm k}$ & $M_{\rm Ni}$ & $M_{\rm ej}$ & $E_{\rm k}/M_{\rm ej}$ & Ref.\\
& [$10^{51}$ erg]& [M$_\odot$]& [M$_\odot$]& [$10^{51}$ erg/M$_\odot$]&\\
\hline
SN 1998bw& 50& 0.38--0.48&11&4.5&2, 3\\
SN 2003dh&40& 0.25--0.45& 8& 5&4 \\
SN 2003lw&60&0.45--0.65&13&4.6&5\\
SN 2006aj&2&0.21&2&1& 6, 7\\
SN 2010bh&  9.7& 0.12& 3.2& 3& 1\\
\hline
SN 2002ap&4&0.1&2.5&1.6&8\\
SN 2003jd&7&0.36&3&2.3&9\\
SN 2009bb & 18 & 0.22 & 4.1 & 4.4 & 10\\
\hline
\end{tabular}
\end{center}
{\footnotesize (1) This paper; (2) \citealt{Iwamoto98bw}; (3) \citealt{Mazzali98bw};  (4) \citealt{Mazzali03dh}; (5) \citealt{Mazzali03lw};
(6) \citealt{Pian06aj}; (7) \citealt{Mazzali06aj}; (8) \citealt{Mazzali02ap}; (9) \citealt{Valenti03jd}; (10) \citealt{Pignata09bb}.}
\end{table*}

\end{document}